\documentclass[acmlarge]{acmart}

\usepackage{epsfig}
\usepackage{epstopdf}
\usepackage{colortbl}
\usepackage{subfigure}
\usepackage{booktabs}
\usepackage{multirow}
\usepackage{soul}
\usepackage{graphicx}      
\usepackage[ruled]{algorithm2e} 

\SetAlFnt{\small}
\pdfoutput=1
\SetAlCapFnt{\small}
\SetAlCapNameFnt{\small}
\SetAlCapHSkip{0pt}
\IncMargin{-\parindent}

\acmJournal{IMWUT}
\acmVolume{0}
\acmNumber{0}
\acmArticle{0}
\acmYear{2017}
\acmMonth{0}
\acmArticleSeq{0}

\setcopyright{acmcopyright}

\acmDOI{0000001.0000001}

\received{XXXX 2017}
\received[accepted]{XXXX 2017}

\newcommand{\new}[1]{\textcolor{red}{{#1}}}
\begin{document}
\title{Smartphone App Usage Prediction Using Points of Interest}

\author{Donghan Yu, Yong Li, Fengli Xu}
\orcid{1234-5678-9012-3456}
\affiliation{
  \institution{Tsinghua University}
  \streetaddress{Tsinghuayuan Rd}
  \city{Beijing}
  \state{Beijing}
  \postcode{100084}
  \country{China}
  }

\author{Pengyu Zhang}
\orcid{1234-5678-9012-3456}
\affiliation{
  \institution{Stanford University}
  \streetaddress{Tsinghuayuan Rd}
  \city{Beijing}
  \state{Beijing}
  \postcode{100084}
  \country{China}
  }    
  
\author{Vassilis Kostakos}
\affiliation{
  \institution{University of Melbourne}
  \department{School of Computing and Information Systems}
  \city{Melbroune}
  \state{VIC}
  \postcode{3010}
  \country{Australia}
  }
  

\begin{abstract}
In this paper we present the first population-level, city-scale analysis of application usage on smartphones. Using deep packet inspection at the network operator level, we obtained a geo-tagged dataset with more than 6 million unique devices that launched more than 10,000 unique applications across the city of Shanghai over one week. We develop a technique that leverages transfer learning to predict which applications are most popular and estimate the whole usage distribution based on the Point of Interest (POI) information of that particular location. We demonstrate that our technique has an $83.0\%$ hitrate in  successfully identifying the top five popular applications, and a $0.15$ RMSE when estimating usage with just $10\%$ sampled sparse data. It outperforms by about $25.7\%$ over the existing state-of-the-art approaches. Our findings pave the way for predicting which apps are relevant to a user given their current location, and which applications are popular where. The implications of our findings are broad: it enables a range of systems to benefit from such timely predictions, including operating systems, network operators, appstores, advertisers, and service providers.
\end{abstract}

%
%


%
%



\thanks{This work is supported by the research fund of Tsinghua University - Tencent Joint Laboratory for Internet Innovation Technology.}

\maketitle


\section{Introduction}

We present the first population-level, city-scale analysis of application usage on smartphones. Our work contributes to the growing body of research that has been spurred by the flourishing appstore economy, and which has motivated researchers in recent years to investigate users' smartphone application usage behaviour. For example, previous work has looked at how individuals download, install, and use different applications on their personal devices \cite{Xu:2013:PCC:2493988.2494333,Srinivasan:2014:MMY:2632048.2632052,Shin:2012:UPM:2370216.2370243}. Typically, such work has investigated behaviour at an individual level, and often attempting to cluster users based on similarities of their behaviours \cite{RN10681}. As such, most studies only have sampled information about application usage, either collected from the mobile devices of volunteers or monitored on the network side with low penetration.

Despite the ubiquity and mobility of smartphones and personal devices, very little work to date has investigated how context, and in particular physical location, affects application usage. For example, some prior work has investigated which applications people use at "home" versus at "work" versus "on the go" \cite{RN10721}. However, such work does not capture the rich urban or socioeconomic characteristics of a location explicitly, but only through the prism of the purpose that a particular location plays in a participant's everyday life.

Understanding how mobile application usage patterns vary across different types of locations in large scale urban environments is extremely valuable for operating systems, profiling tools, appstores, service providers, and even city managers. For example, appstores can promote different types of apps based on the location of the user, and operating systems or profiling tools can provide shortcuts to the apps most likely to be used at the current location. A strength of our work is that our model can only rely on the list of nearby Points of Interest (POIs) at any given location. This means that actual GPS coordinates do not have to be disclosed, thus ensuring a certain level of privacy. Additionally, our model is highly extensible and can also make predictions using other types of data, such as anonymized user identification list of each location.

Our analysis investigates the rich relationship between the characteristics of a physical location and the smartphone apps that people use at that location. Specifically, we consider the urban characteristics of a location as reflected by its patterns of socio-economic activity, infrastructure, and social cohesion. To achieve this, we analyse the \emph{types} and \emph{density} of POIs at any given location, and correlate them to the popularity of various applications at that location. Intuitively, we hypothesize that at certain types of locations users are more likely to exhibit stronger interest in a particular class of applications. For example, a region containing a number of universities and schools has a high probability to be an educational area, and we expect that people may be more likely to use educational applications on their devices. On the other hand, a region usually contains a variety of POIs, resulting into different app usage patterns. Thus, we hypothesize that we are able to use publicly available POI data to predict the app usage in each type of location. This requires overcoming the challenge of merging both POI and app usage datasets, and developing predictive techniques that take advantage of both datasets.

In this paper, we propose a novel transfer learning technique to predict the smartphone application usage at any given location by considering the POIs in that location. We analysed a large scale application usage dataset with more than 6~million unique devices launching more than 10,000 unique applications covered by over 9,800 base station sectors. The data was collected from the mobile network of Shanghai over a period of one week. Our analysis investigates the challenges and opportunities of fusing POI data with application usage records to estimate the application usage in each area of the city. The contribution of our work is three-fold:
\begin{itemize}
  \item We are the first to propose and investigate the idea of using publicly available POI data to help predict and estimate application usage at a given location. A key contribution is that with our work, researchers can simply rely on easy-to-get POI data for estimating application usage at a given location, without having to collect hard-to-get application usage data from users' devices.
  \item We propose a transfer learning method based on collaborative filtering to estimate application usage. Our method transfers the knowledge domain of POIs and users into the domain of application usage by uncovering and learning the underlying latent correlations between these domains. Moreover, we incorporate temporal dynamics into our model to achieve high prediction accuracy. Our proposed method is computationally efficient in achieving knowledge transfer between domains.
  \item We evaluate the performance of our proposed system and compare it against state-of-the-art baseline predictors. Our evaluation considers a variety of scenarios and parameters. The results demonstrate that our technique can reliably learn POI information to help predict application usage. It achieves $83.0\%$ hitrate in predicting the top five popular applications at a location, and a $0.15$ RMSE when estimating the total usage distribution, thus improving by $25.7\%$ over state-of-the-art approaches.
 \end{itemize}

\section{Data}

\subsection{Smartphone Network Traces}
Our dataset contains anonymized cellular data accessing traces obtained by Deep Packet Inspection (DPI) appliances. Data was collected from mobile cellular network in Shanghai, one of the major metropolitan areas in China. Data requests on the mobile network were passively inspected and captured the identification (ID) of each mobile device (anonymized), ID and location of the base station sector(s) from which the request was made, start and end timestamps of the data connection. In addition, DPI revealed the HTTP request or reponse URL with path and parameters, visited domain and user-agent field of the client.

Using the captured data, we can infer the smartphone app that is likely to have generated these requests. Because many apps make Internet requests, for example checking for new versions or upload data, we were able to inspect and identify the particular HTTP headers that any given app uses. Note that our approach has an inherent limitation: it does not capture smartphone apps that make absolutely no network requests, nor apps that make requests solely through WiFi networks. Thus, apps that do not use cellular networks are excluded from our analysis. However, a recent  report\footnote{https://techcrunch.com/2017/05/04/report-smartphone-owners-are-using-9-apps-per-day-30-per-month/} claims that the daily average number of apps used by smartphone users in China is about 11, which very similar to the daily average number 9.2 of apps obtained by each user in our DPI dataset. This indicates that the number of apps that do not request networks is non-trivial but negligible.

\begin{figure}
\centering
\subfigure[Time interval between two consequetive records]{\includegraphics[width=.45\textwidth]{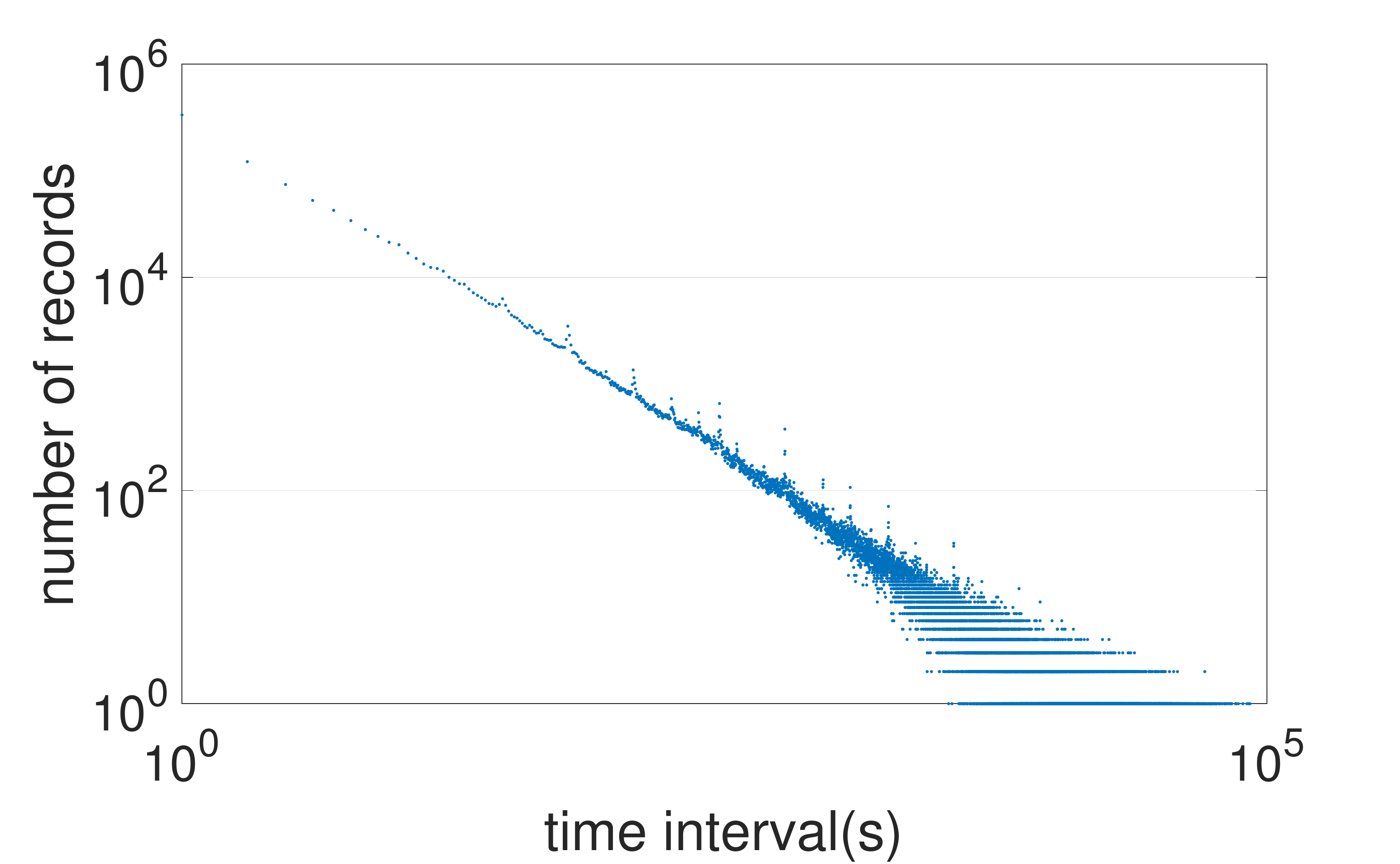}}
\subfigure[Number of records for each device]{\includegraphics[width=.45\textwidth]{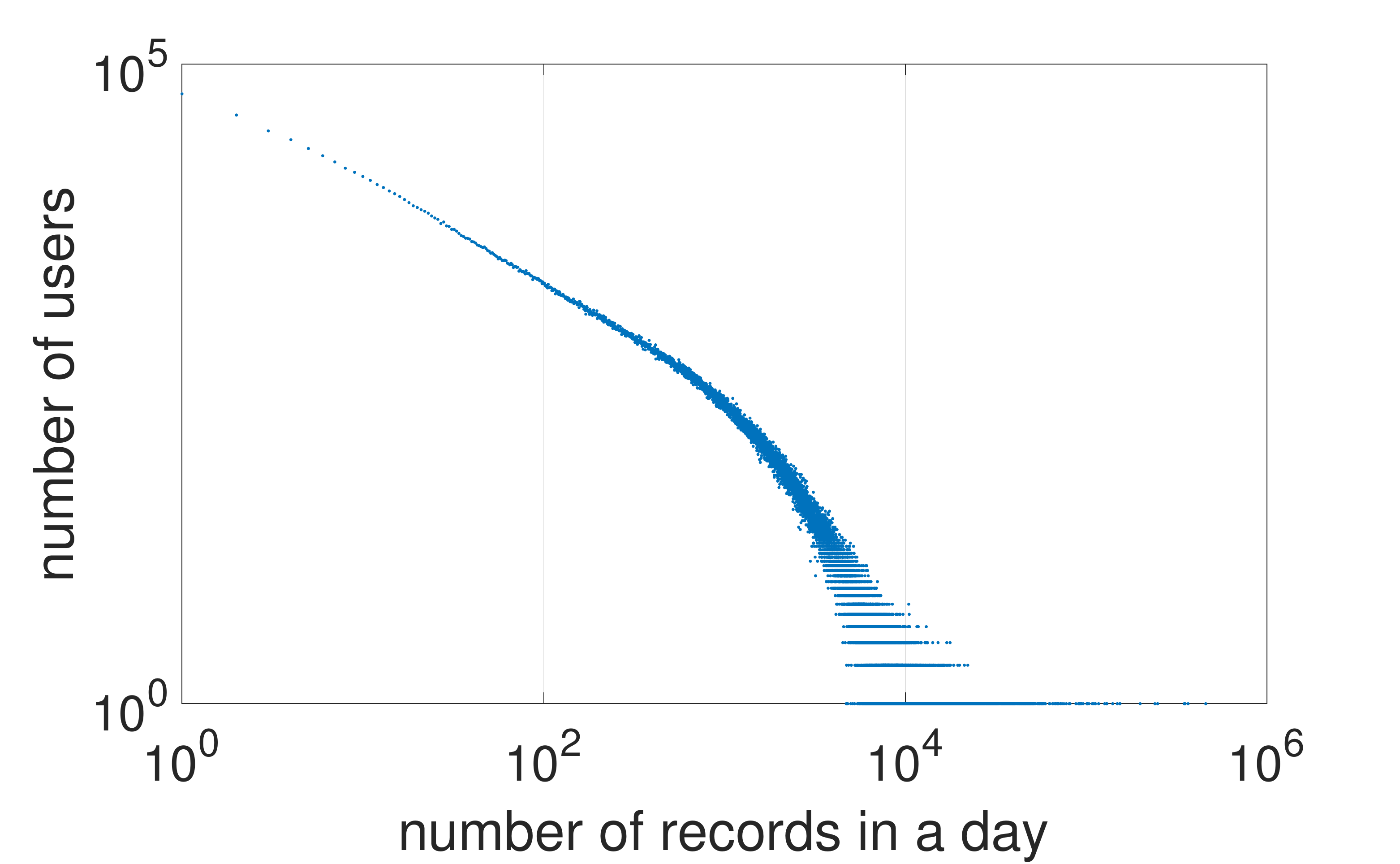}}
\caption{The dataset exhibits fine-grain properties.} \label{fig:dat_chr}
\end{figure}

Overall, the trace dataset contains over 6 million unique devices, 10000 unique applications, and 9800 base station sectors. It spans a period of 7 days. An app-usage record gets created after every network request with the granularity of every packets sent from the mobile device. The average time interval between two consecutive records for a given device is 222 seconds. In Fig.~\ref{fig:dat_chr}(a), we plot the interval between two records against the frequency of observation for that interval. It reveals a power law distribution with most measurements below 1000 seconds. Similarly, in Fig.~\ref{fig:dat_chr}(b) we plot the number of daily records for any given user against the frequency of observation. It reveals a power law distribution with exponential cut-off. As such, the number of records generated by a user each day scales smoothly between the range of 1 to 1000, but drops drastically after 1000. The most active mobile user can generate up to hundreds of thousands of records on a given day. 

It is worth pointing out that privacy issues of this dataset are carefully considered and measures are taken to protect the privacy of these mobile users. Our dataset is collected via a collaboration with the mobile network operator, and the data does not contain any personally identifiable information. The ``user ID'' field has been anonymized (as a bit string) and does not contain any user meta-data. All the researchers are regulated by strict non-disclosure agreement and the dataset is located in a secure off-line server. In our dataset, more than 95\% of traffic uses HTTP at the time of data collection. Even though certain apps may use HTTPS protocols, they typically also have some part of their traffic use plain HTTP, thus providing us the opportunity to infer the identity of the application.

\subsection{Inferring Application Identity}

To establish ground truth in our dataset, we need to reliably infer the identity of the application that has made the network requests we captured. In the HTTP header captured by our DPI, various fields are utilized as the identifiers of the apps to communicate with their host servers or third party services. The hosting servers need to distinguish between different applications in order to provide appropriate content. Therefore, we are able to identify the app making a network request by inspecting those HTTP header identifiers. We utilized a systematic framework for classifying network traffic generated by mobile applications: SAMPLES \cite{SAMPLES}. It uses constructs of conjunctive rules against the application identifier found in a snippet of the HTTP header. The framework operates in an automated fashion through a supervised methodology over a set of labeled data streams. It has been shown to identify over 90\% of these applications with 99\% accuracy on average \cite{SAMPLES}, and we manually verified its accuracy for a small subset of records. This method enabled us to accurately label most of the applications found in our dataset. To present a more clear view of our application dataset, we list the two most popular apps in each app category in Table~\ref{tab:app_ctg}.

\begin{table*}
\newcommand{\tabincell}[2]{\begin{tabular}{@{}#1@{}}#2\end{tabular}}  
\begin{center}
\begin{tabular}{c|c|c||c|c|c}
\hline
\textbf{$i$} & \textbf{$i^{th}$ category} & \textbf{Top2 ranked app} & \textbf{$i$} & \textbf{$i^{th}$ category} & \textbf{Top2 ranked app} \\
\hline
1 & Games & \tabincell{c}{KaiXinXiaoXiaoLe, HuanLeDouDiZhu} & 2 & Videos & \tabincell{c}{iQiyi, QQLive} \\
\hline
3 & News &  \tabincell{c}{QQNews,JinRiTouTiao} & 4 & Social &  \tabincell{c}{QQ, Wechat}\\
\hline
5 & E-Commerce &  \tabincell{c}{Taobao, JingDong} & 6 & Finance &  \tabincell{c}{TongHuaShun, ZiXuanGu} \\
\hline
7 & Real Estate &  \tabincell{c}{LianJia, AnJuKe} & 8 & Travel &  \tabincell{c}{ctrip, QuNaEr} \\
\hline
9 & Life Service &  \tabincell{c}{Meituan, DaZhongDianPing} & 10 & Education &  \tabincell{c}{YouDao Dictionary, ZhiHu} \\
\hline
11 & Taxi &  \tabincell{c}{DidiTaxi, DiDaPinChe} & 12 & Music &  \tabincell{c}{QQMusic, AJiMiDeFM} \\
\hline
13 & Map &  \tabincell{c}{GaoDeMap, BaiduMap} & 14 & Reading &  \tabincell{c}{QQReader, ZhuiShuShenQi} \\
\hline
15 & Fashion &  \tabincell{c}{MoGuJie, MeiTuXiuXiu} & 16 & Office &  \tabincell{c}{189Mail, QQMail} \\
\hline
\end{tabular}\caption{Top2 ranked apps for 16 app categories}\label{tab:app_ctg}
\end{center}
\end{table*}

\subsection{Points of Interest}

Intuitively, our approach considers base station sectors as landmarks that reveal the location of the user when their smartphone made a particular network request. In turn we can identify the nearby Points of Interest from existing open datasets, and use them to provide additional context.

We utilize Voronoi diagrams \cite{aurenhammer1991voronoi} to partition the city and obtain the coverage area of each base station sector and its "nearby" POIs. Specifically, the Voronoi diagram partitions the coverage area for each base station as $\{b(l_1), b(l_2), ..., b(l_M )\}$, where any POI $p_i \in b(l_i)$ satisfies that for any POI $l_j \not= l_i$, the Euclidean distance between $p_i$ and $l_i$ is smaller than that between $p_i$ and $l_j$. In this manner, we built the Voronoi polygons based on the spatial location of base station sectors.

POIs can be considered as indicators associated with specific urban and economic functions such as shopping, education, or entertainment. As such, POIs characterises the socioeconomic function of a location served by a particular base station sector. In our analysis, we obtained all the POIs (about 750,000 items) of Shanghai city from BaiduMap (one of largest POI databases in China). They are classified as 17 types, including Food, Hotel, Shopping, Life Service, Beauty, Tourism, Entertainment, Sports, Education, Culture Media, Medical Care, Automotive Service, Traffic Facilities, Finance, Real Estate, Company and Government. In our system, this POI dataset is utilized as the input to predict the application usage.

\section{Analysis}

\subsection{Conceptual approach}


Intuitively, we hypothesize that POI information represents the attributes of a location, and we argue that such attributes have important impact on the types of apps that people use. For example, we argue that near tourist attractions, people are less likely to use office-type apps such as WPS and Email, and more likely to use photo apps or travel apps. 

\begin{figure}[!htp]
\centering
\subfigure[For different types of locations (i.e. locations with relatively higher frequency of certain types of POIs), we calculate the relative popularity of application categories.]{
\label{Fig.lableapptype}
\includegraphics[height=5cm, width = 8cm]{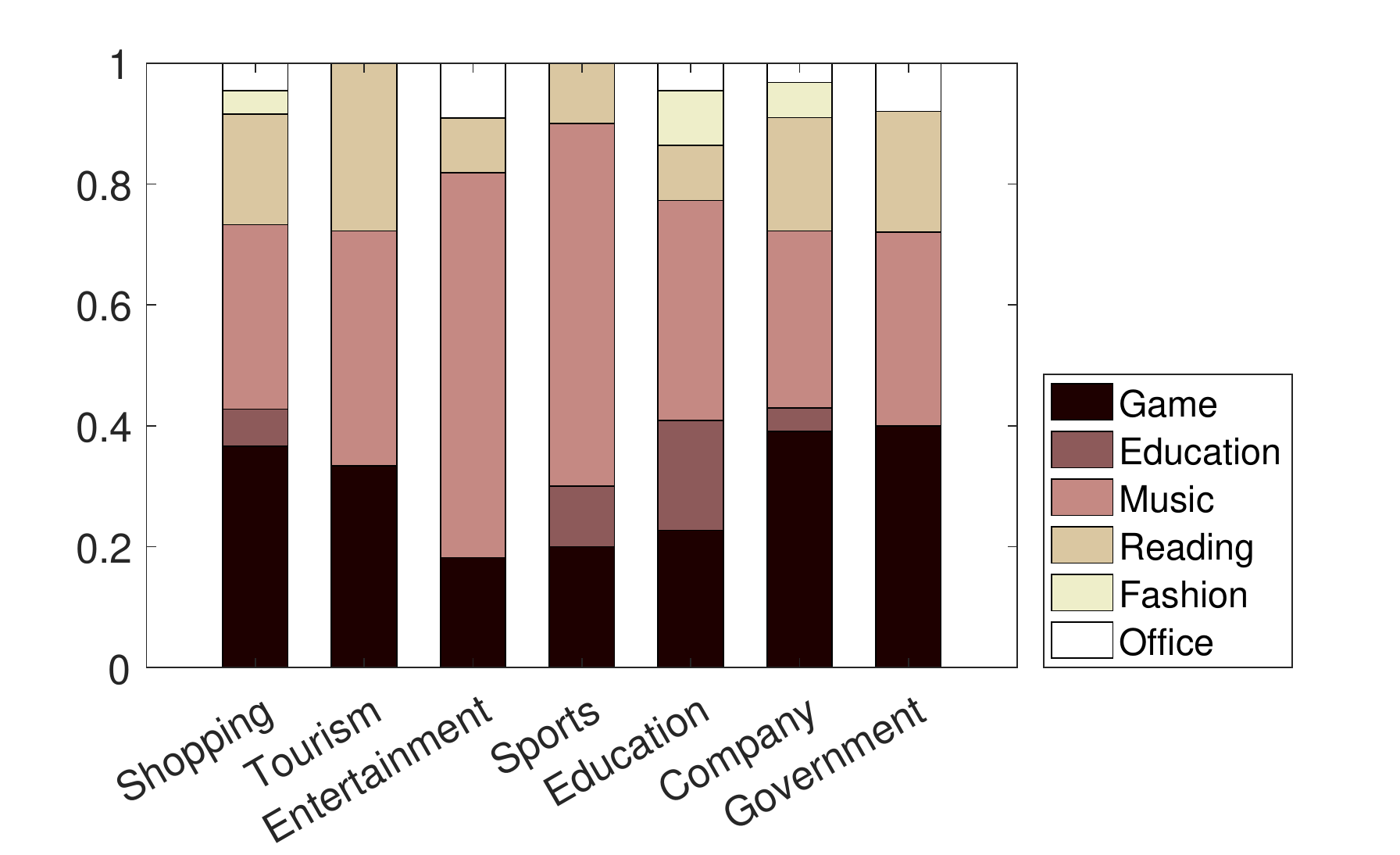}}
\subfigure[Cumulative Distribution Function of the statistical correlation between the vectors of app usage and POI information.]{ 
\label{Fig.lablestatistical}
\includegraphics[height=5cm]{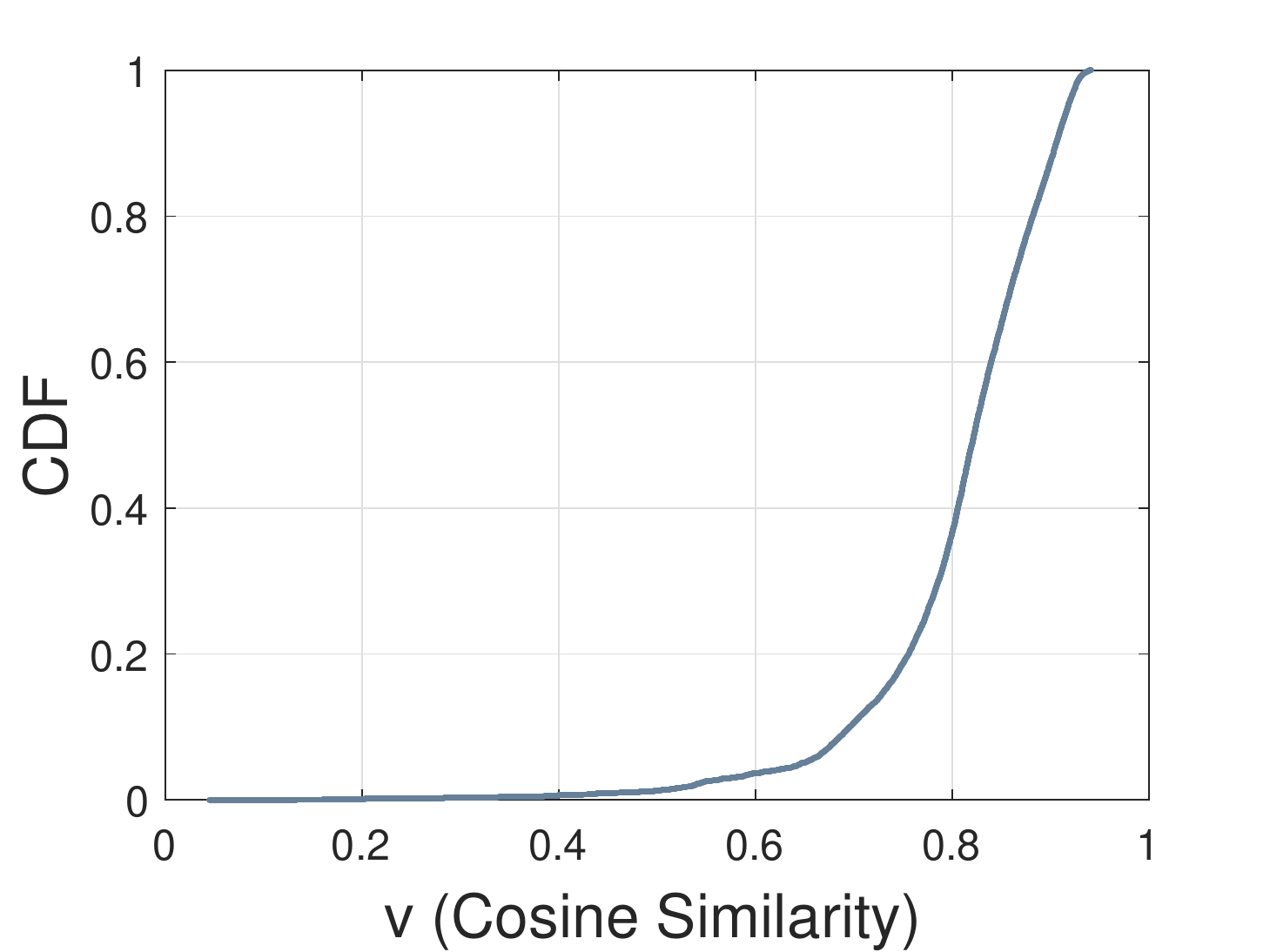}}
\caption{Intuitive and statistic correlation between app usage and POI information}
\label{Fig.label.parameters}
\end{figure}

Hence, our analysis focuses on using our real word dataset to investigate the relationship between the location where people use apps on their smartphone, and the nearby POIs at these locations. We cluster all locations according to their most popular POI category, which means that the locations in the same cluster share the same most popular type of POI. For instance, we label a location as \emph{X-Location} if this location's most popular POIs are of type $X$, which can be e.g. Hotels. Then, we sum up the app usage of locations in each cluster respectively. Their distribution in terms of different app types are partly illustrated in Fig.~\ref{Fig.lableapptype}. It can be observed that at \emph{Tourism-Locations} and \emph{Sports-Locations}, people are less likely to use Fashion or Office apps. Music apps are used more frequently at the \emph{Sports-Locations} and people tend to use Education apps more often at \emph{Education-Locations}.



To further quantify the relationship between POIs and apps, suppose the total number of locations is $m$, then we define a location-app-corr matrix $M$ and location-POI-corr matrix $N$, which satisfy that $M_{ij}$ denote the Cosine Similarity between location $i$ and $j$ based on POI, and $N_{ij}$ denote the Cosine Similarity between location $i$ and $j$ based on app usage data (measured as the total number records in our dataset). More precisely, we denote the app usage vector of location $i$ as $\textbf{r}_i= [r_{i1}, r_{i2}, \ldots, r_{in}]$ for $n$ apps, where each $r_{ij}$ is the number of times (i.e. records in our dataset) the app $j$ is used at location $i$, and POI information vector of location $i$ as $\textbf{q}_i = [q_{i1}, q_{i2}, \ldots, q_{il}]$ for $l$ types of POIs. Then we have:
\begin{eqnarray}
M_{ij} = cos(\textbf{r}_i, \textbf{r}_j), N_{ij} = cos(\textbf{q}_i, \textbf{q}_j),\forall i,j=1,\ldots,m \ .
\end{eqnarray}

We define $\textbf{v} = [v_{1}, v_{2}, \ldots, v_{m}]$, where $v_{i}$ is the Cosine Similarity between app usage and POI information at location $i$, as:
\begin{eqnarray}
v_{i} = cos (\textbf{m}_i, \textbf{n}_i), \forall i=1,\ldots,m \ ,
\end{eqnarray}  
where $\textbf{m}_i$ and $\textbf{n}_i$ denote the $i$th rows of matrix $M$ and $N$ respectively. The Cumulative Distribution Function (CDF) of $\textbf{v}$ is shown in Fig.~\ref{Fig.lablestatistical}. This plot indicates that for nearly half of the locations (i.e. above $40\%$), the app usage and POI information are strongly correlated (above $0.85$). All these results indicate the strong correlation between the App usage and POI information, which demonstrates the feasibility of our idea that transferring the knowledge of POI data helps predict app usage.


\subsection{Predicting app usage}
Now, we show how to use POI information to predict the app usage at any given location. For a location $i$, we have a $n$-dimensional count vector $\textbf{r}_i= [r_{i1}, r_{i2}, \ldots, r_{in}]$ for $n$ apps. However, the frequency of app usage at one location may be distributed in a large range due to the power law distribution we have observed, thus preprocessing is necessary for better transferring and preventing overfitting. We denote the location-app matrix as $X_{m\times n}$, and define its entries as:
\begin{eqnarray}
X_{ij} = \log(r_{ij})/\max(\log{(r_{ij}))} , \forall i=1,\ldots,m;j=1,\ldots,n.
\end{eqnarray}
We divide the entries by $\max(\log{(r_{ij})})$ to make sure the range set within $[0, 1]$. When $X_{ij}$ is missing, it indicates there is no observation of app $j$ being used at location $i$. This is conceptually different from meaning that there is no possibility to use the app at location $i$.

Next,  we calculate the profile of each base station by considering the category and density of nearby POIs. Denote the count vector for a location $i$ as $\textbf{q}_i = [q_{i1}, q_{i2}, \ldots, q_{il}]$ for $l$ types of POIs. Consider that some types of POIs (e.g. hotels) are more popular than others (e.g. tourist attractions), we further normalize these counts using the metric term-frequency inversed-document-frequency (TF-IDF) \cite{steyvers2007probabilistic}, which is designed to reflect how important a word is to a given document, to obtain a location-POI matrix $Y_{m\times l}$. More precisely, we have each entry of $Y$ as follows,
\begin{eqnarray}
Y_{ip} = \frac{q_{ip}}{\sum_{p=1}^lq_{ip}}\cdot \log\frac{|\{\textbf{q}_i\}|}{|\{\textbf{q}_i:q_{ip}>0\}|} , \forall i=1,\ldots,m;p=1,\ldots,l \ ,
\end{eqnarray}
where $|\textbf{q}_i|$ is the number of all the count vectors (i.e. number of locations), and $|\{\textbf{q}_i:q_{ip}>0\}|$ is the number of count vectors (i.e. locations) having non-zero $p$-th type of POIs. Using this processing method, we can increase the weights for these important POIs that are fewer but unique (e.g. tourist attractions), and decrease the weights for the POIs that may be extensively distributed across a city (e.g. hotels).

To make our prediction model more accurate, we take personal preference into consideration, since we also know the anonymized "user ID" set of each location. One approach could consider the "user ID" at each location, and construct a location-user matrix for collective matrix factorization. However, this method is rather cumbersome and does not scale due to millions of users in total. Instead, we construct a user-based location correlation matrix $Z$. We denote the "user ID" set of location $i$ as $U_i$, and the correlation between location $i$ and $j$ is calculated as follows:

\begin{eqnarray}
Z_{ij} = \frac{2|U_i \bigcap U_j|}{|U_i| + |U_j|} , \forall i=1,\ldots,l;j=1,\ldots,l \ ,
\end{eqnarray}
where $\bigcap$ means the intersection of two sets and $| \ |$ gets the cardinality  of a set. From the equation, we observe a higher correlation between two locations that share more common users.

After calculating the location-app matrix $X$, location-POI matrix $Y$ and location correlation matrix $Z$, we use transfer learning to find a latent feature representation for locations, apps and POIs\footnote{Our transfer learning model with its generative model, is described in detail in the Appendix \uppercase\expandafter{\romannumeral1}}. What we transfer among the location-POI domain, location-user domain and location-app domain is the latent feature of locations.  We denote $L \in R^{K\times m}$, $A \in R^{K\times n}$ and $P \in R^{K\times l}$ to represent the latent location, app and POI matrices respectively, with column vectors $\textbf{l}_i$, $\textbf{a}_j$, $\textbf{p}_k$ representing the $K$-dimensional location-specific latent feature vector of location $i$, app-specific latent feature vector of app $j$, and POI-specific latent feature vector of POI $k$, respectively. For location latent feature vector $\textbf{l}_i$, we consider its components as functionality-based feature $\textbf{l}_i^1$ and user preference-based feature $\textbf{l}_i^2$, which means that $\textbf{l}_i = \textbf{l}_i^1 + \textbf{l}_i^2$. The corresponding location feature matrices are $L_1$ and $L_2$.

We denote the flag matrix as $I$ for location-app data. If the usage data of app $j$ at location $i$ is known, then $I(i,j) = 1$, otherwise $I(i,j) = 0$. Maximizing the log-posterior over the latent feature of locations, apps and POIs is equivalent to minimizing the following objective function, which is a sum of squared errors with quadratic regularization terms as follows,

\begin{eqnarray}\label{cmf}
\begin{split}
\zeta (L_1,L_2,A,P)  = & \frac{1}{2}|| I \circ \left(X - g\left((L_1 + L_2)^\top A\right) \right)||_F^2 +\frac{\alpha}{2}|| Y - g\left(L_1^\top P\right) ||_F^2 +\frac{\beta}{2}||  Z - g\left(L_2^\top L_2\right) ||_F^2\\
& +  \left( 
\frac{\lambda_l^1}{2}||L_1||_F^2 +
\frac{\lambda_l^2}{2}||L_2||_F^2+
\frac{\lambda_a}{2}||A||_F^2 + \frac{\lambda_p}{2}||P||_F^2
\right) \ ,
\end{split}
\end{eqnarray}

where $\circ$ means the point-wise matrix multiplication and function $g(x)$ is the point-wise logistic function $g(x) = 1/(1+exp(-x))$ to bound the range within $[0,1]$. $\alpha$ is the weight of location-POI data we use for transfer learning, $\beta$ means the weight of user-based location correlation data.

Finally, we take temporal dynamics into consideration. Since users' app usage varies with time, the statistical usage in any one location is also time-varying. For example, people tend to use office apps during work hours. This suggests that time-of-day is likely to be an important factor in determining which applications people are using. As such, shorter time periods are likely to be more "homogeneous" rather than longer ones. Since collaborative filtering is based on an assumption of homogeneity, the method makes better predictions for narrower time frames. To incorporate time into our model, we consider the final loss function $\zeta$ is the sum of time-specific loss function of $\zeta (L_{1,t},L_{2,t},A,P_t)$ (\ref{time-1}) during different time periods plus the regularization based on temporal continuity of time-specific latent feature vectors (\ref{time-2}). Note that the static location-POI matrix $Y$ does not change, and we share the same app latent feature $A$ at different time-specific loss functions to transfer knowledge among them since we consider the latent feature of apps keeps static, while the latent feature of POI is time-varying as Bromley et al. \cite{bromley2003disaggregating} noted that the effects of POI vary substantially during the day. It can be expressed as follows,

\begin{eqnarray}\label{time-1}
\begin{split}
\zeta (L_{1,t}.L_{2,t},A,P_t)  = &
 \frac{1}{2}|| I_t\circ \left(X_t - g\left((L_{1,t} + L_{2,t})^\top A\right) \right)||_F^2 +\frac{\alpha}{2}||  Y - g\left(L_{1,t}^\top P_t\right) ||_F^2 +\frac{\beta}{2}||  Z_t - g\left(L_{2,t}^\top L_{2,t}\right)||_F^2  \\
 &  + \left( 
\frac{\lambda_l^1}{2}||L_{1,t}||_F^2 +
\frac{\lambda_l^2}{2}||L_{2,t}||_F^2 +
\frac{\lambda_a}{2}||A||_F^2 + \frac{\lambda_p}{2}||P_t||_F^2
\right),
\end{split}
\end{eqnarray}

\begin{eqnarray}\label{time-2}
\begin{split}
\zeta = \sum_t \zeta (L_{1,t}.L_{2,t},A,P_t) + \left(\frac{\lambda_1}{2}  \sum_t\left(||L_{1,t} - L_{1,t-1}||_F^2 +
||L_{2,t} - L_{2,t-1}||_F^2\right)
 + \frac{\lambda_2}{2} \sum_t||P_t - P_{t-1}||_F^2 \right).
\end{split}
\end{eqnarray}

There exist several methods to reduce the time complexity of model training, and we adopted mini-batch gradient descent approach to learn the parameters. With random sampling, the cost of the gradient update no longer grows linearly in the number of entities related to latent feature vectors, but only in the number of entities sampled. The hyper-parameters, i.e., number of latent features and regularization coefficient, are set by cross-validation. After learning the latent features of locations and apps, we can reconstruct the location-app matrix during different time periods or of the whole time period.


\section{Evaluation}
To evaluate the accuracy of our predictive model, we conduct extensive experiments to compare the prediction capabilities of our proposed model against multiple alternative models. Our evaluation aims to answer the following questions:

1) How accurate is our prediction model at different sparsity levels?

2) What is the impact of the number of users on the model's performance?

3) How does the space granularity affect the model's performance?

4) How does our model perform for locations for which we do not have any prior app usage data?

Since we have ground truth data available, we adopt three evaluation metrics \footnote{Described in detail in the Appendix \uppercase\expandafter{\romannumeral2}}: Top-$N$ hit rate, Top-$N$ prediction accuracy and Root Mean Square Error (RMSE) to fully evaluate our technique. The first metric is the percentage of locations whose top-$N$ apps are successfully predicted (correct for at least one). This metric is often used for recommender systems, because such systems typically recommend a list of items and expect users to click at least one of them. The second metric reflects the average accuracy on top-$N$ predictions at all locations. Finally, RMSE measures the error between the true and estimated app usage distribution. 

\subsection{Experiments and Baselines Setting}

In our analysis, we use apps that cover most of the records of the whole dataset as the grand truth, which retains about $90\%$ of our records. For time period, we construct 24 time periods by merging the same hour of the seven days into one time period, such as 17:00-18:00. We compare our technique's performance against four baseline approaches: app-only prediction (AOP), Multiple logistic regression (MLR) \cite{walker1967estimation}, single matrix factorization (SMF) \cite{srebro2003weighted} and collective matrix factorization (CMF) \cite{singh2008relational}, which are introduced as follows.

AOP sorts the apps based on the total usage amount of all locations in the training data, and makes predictions based on the sorted app list. 

MLR is a typical machine learning method that estimates the usage of each app independently. For example, assuming the usage of app $j$ in location $I=\{i_1, i_2,...,i_M\}$ is known, we use the POI information vectors of location $I$ as input and the usage of app $j$ in location $I$ as output to train the multiple logistic regression. Then, we predict the usage of app $j$ in other locations using their POI information. The motivation for employing this baseline is to demonstrate the effectiveness of our proposed method to utilize both location-app data and other information for collaborative filtering.

SMF only uses location-app matrix for factorization, ignoring POI information. In particular, this method is equivalent to the case that our loss function (\ref{cmf}) sets $\alpha =\beta= 0$. We employ this baseline to show that with limited number of location-app data (and thus sparse location-app matrix), the prediction results are not good enough. This validates our intuition to use POI and user information to improve the prediction accuracy. 

CMF uses the loss function in (\ref{cmf}), which collectively factorizes the location-app matrix, location-POI matrix and location-correlation matrix, without utilizing the time information. We employ CMF as baselines to mainly justify the usefulness of time information.

Finally, we exclude the top 30 popular apps, e.g. QQ and Wechat, since they are very popular across almost all locations and easy to predict by just choosing the most popular apps (AOP). To validate our rationale, we randomly select $10\%$ of location-app data as training data to test the remaining $90\%$, and choose Top5 hit-rate as the evaluation metric. As shown in Table~\ref{tab:appendix}, we can observe that the result varies according to the number of removed apps, especially when the number is very small. The reason is that there exist many popular apps in our dataset. If we only exclude, say, 10 apps, the AOP method achieves $91\%$ hit-rate with only $10\%$ training data, which indicates good results and there is no need to utilize other more intelligent approaches. Moreover, there are thousands of apps in our dataset, so deleting top 30 does not affect the completeness of our model.

\begin{table}[H]
\begin{center}
\begin{tabular}{c|c|c|c|c}
\hline
\textbf{Number of Excluded Apps} & \textbf{0} & \textbf{10} & \textbf{20}  & \textbf{30} \\
\hline
AOP & $94\%$ & $91\%$ & $78\%$ & $65\%$ \\
\hline
Our Model & $99\%$ & $98\%$ & $91\%$ & $83\%$ \\
\hline
\end{tabular}\caption{Top5 hit-rate of AOP method and Our Model under different numbers of excluded apps} \label{tab:appendix}
\end{center}
\end{table}

\subsection{Effect of Parameters}

In our model, parameters $\alpha$ and $\beta$ controls the contribution of the location-POI and location-user information to the loss function (\ref{time-1}) respectively, and parameter $K$ is the number of latent features (i.e. length of latent vector) in our model. To study the impact of these parameters, we randomly select $20\%$ of location-app data as training data.
To explore the impact of location-POI information, we vary the value of $\alpha$ with fixed $\beta=1$ and plot our model's performance in Fig.~\ref{Fig.label.parameters}(a). The results show that our model's performance first increases and later decreases as $\alpha$ increases. This is because when $\alpha$ is too small, the model cannot fully utilize POI information to capture the location's urban functionality and the relationships among the locations. When $\alpha$ is too large, the POI information dominates the loss function, thus overwhelming the app usage information and user information. With $\alpha\approx5$, our system balances location-POI information and other data well, which achieves the best performance.

\begin{figure}[]
\centering
\subfigure[Top5 hitrate for different $\alpha$ values]{
\label{Fig.sub.1}
\includegraphics[width=5.1cm]{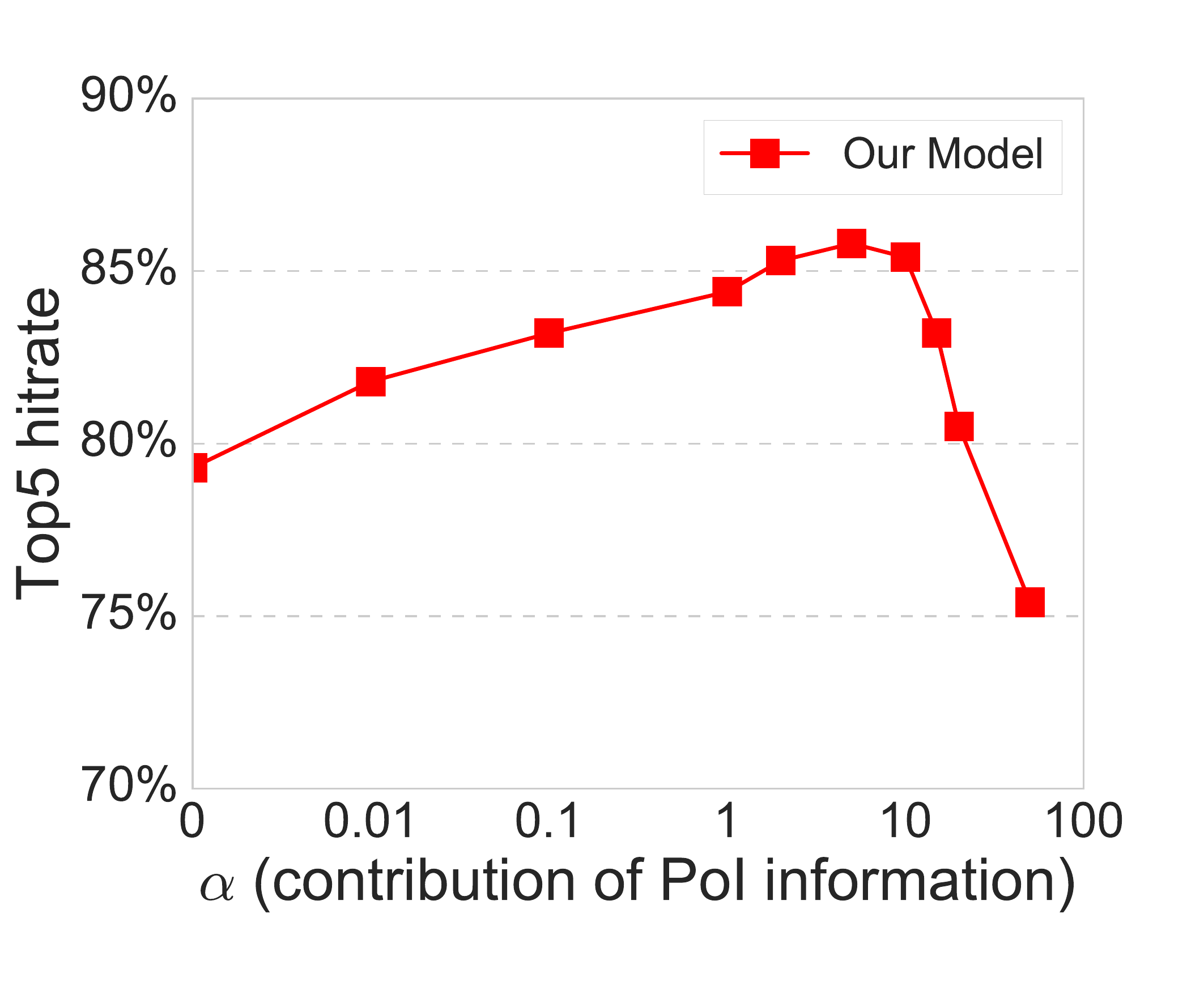}}
\subfigure[Top5 hitrate for different $\beta$ values]{ 
\label{Fig.sub.2}
\includegraphics[width=5.1cm]{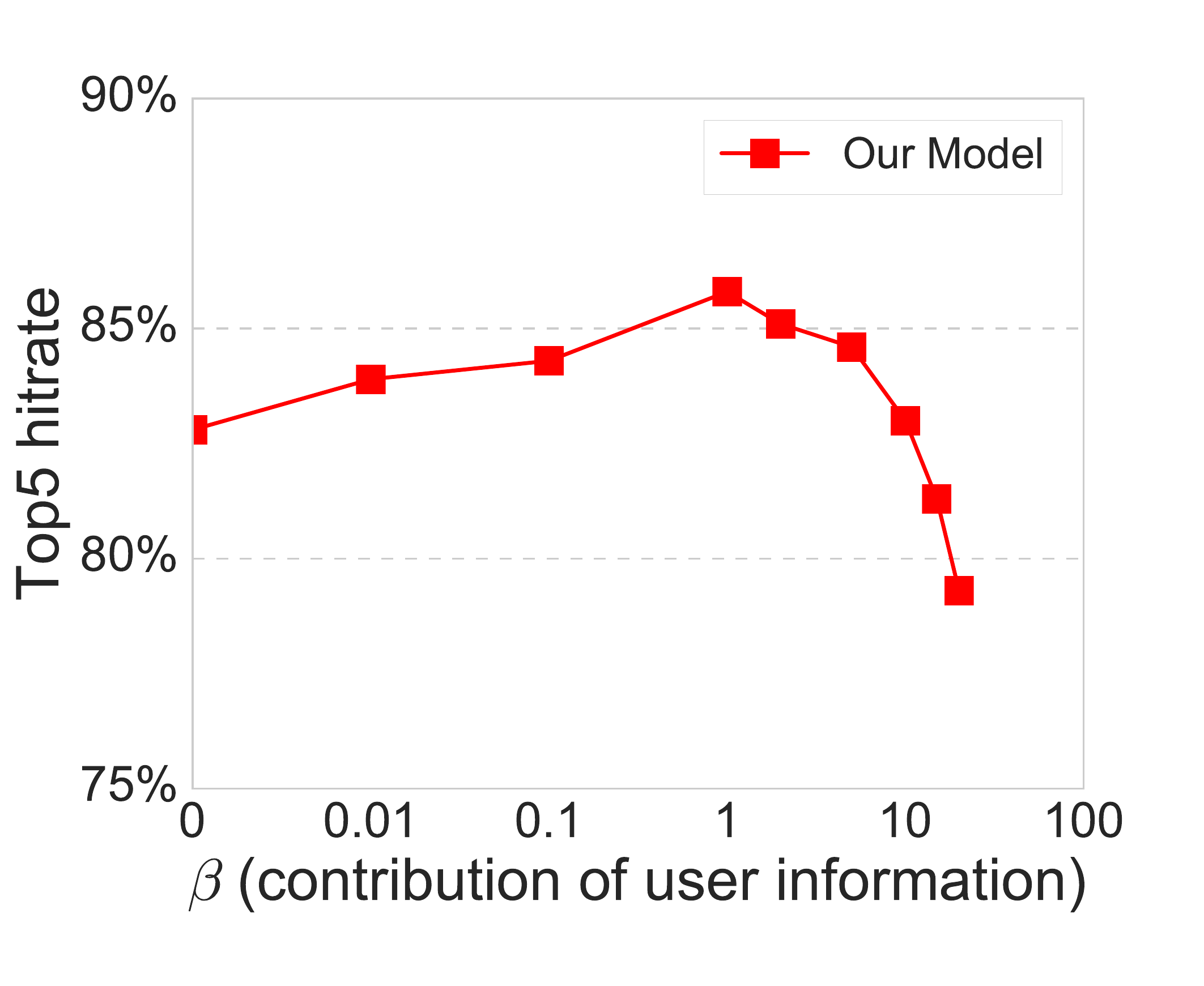}}
\subfigure[Top5 hitrate for different $K$ values]{ 
\label{Fig.sub.3}
\includegraphics[width=5.1cm]{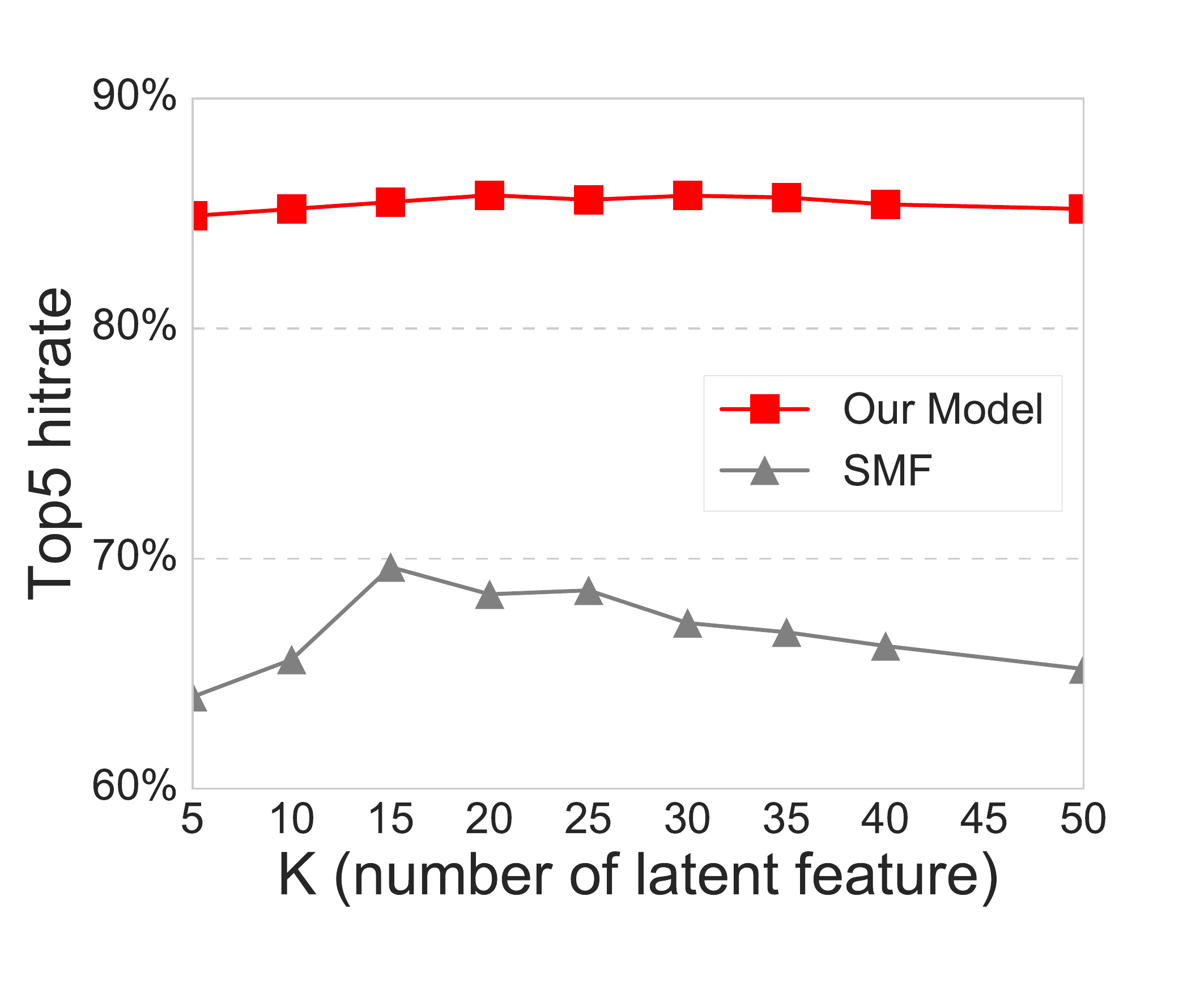}}
\caption{Evaluation metric under different parameter values.}
\label{Fig.label.parameters}
\end{figure} 

We also study the impact of location and user information, by varying the value of $\beta$ and plotting the performance. In this analysis, we fix $\alpha=5$. As is shown in Fig.~\ref{Fig.label.parameters}(b), similarly we observe that our method's hitrate first increases and eventually decreases as $\beta$ grows. When $\beta$ is too small, the user information cannot contribute much to the loss function. When $\beta$ is too large, the user information dominates the loss function, so that prediction will be made without fully considering other factors. With $\beta \approx 1$, our system achieves best performance. In this case, the top-5 hit rate achieves $85.8\%$, which means we can successfully complete the recommendation at about $86\%$ locations.
 
In terms of parameter $K$ (i.e., number of latent features in our model), intuitively, increasing it could add more flexibility to the model. However, after reaching the peak, further increasing $K$ degrades the performance, which may be caused by overfitting with redundant parameters. The results in Fig.~\ref{Fig.label.parameters}(c) for different values of $K$ show that our proposed method performs equally well under varying $K$ values, while SMF is not stable and changes the hit rate by $6\%$ (from $70\%$ to $64\%$). These results indicate the strong robustness of our model.

\subsection{Effect of Varying Sparsity}

\begin{figure}[]
\centering
\subfigure[Top5 hitrate]{
\label{Fig.sub.1}
\includegraphics[width=5.1cm]{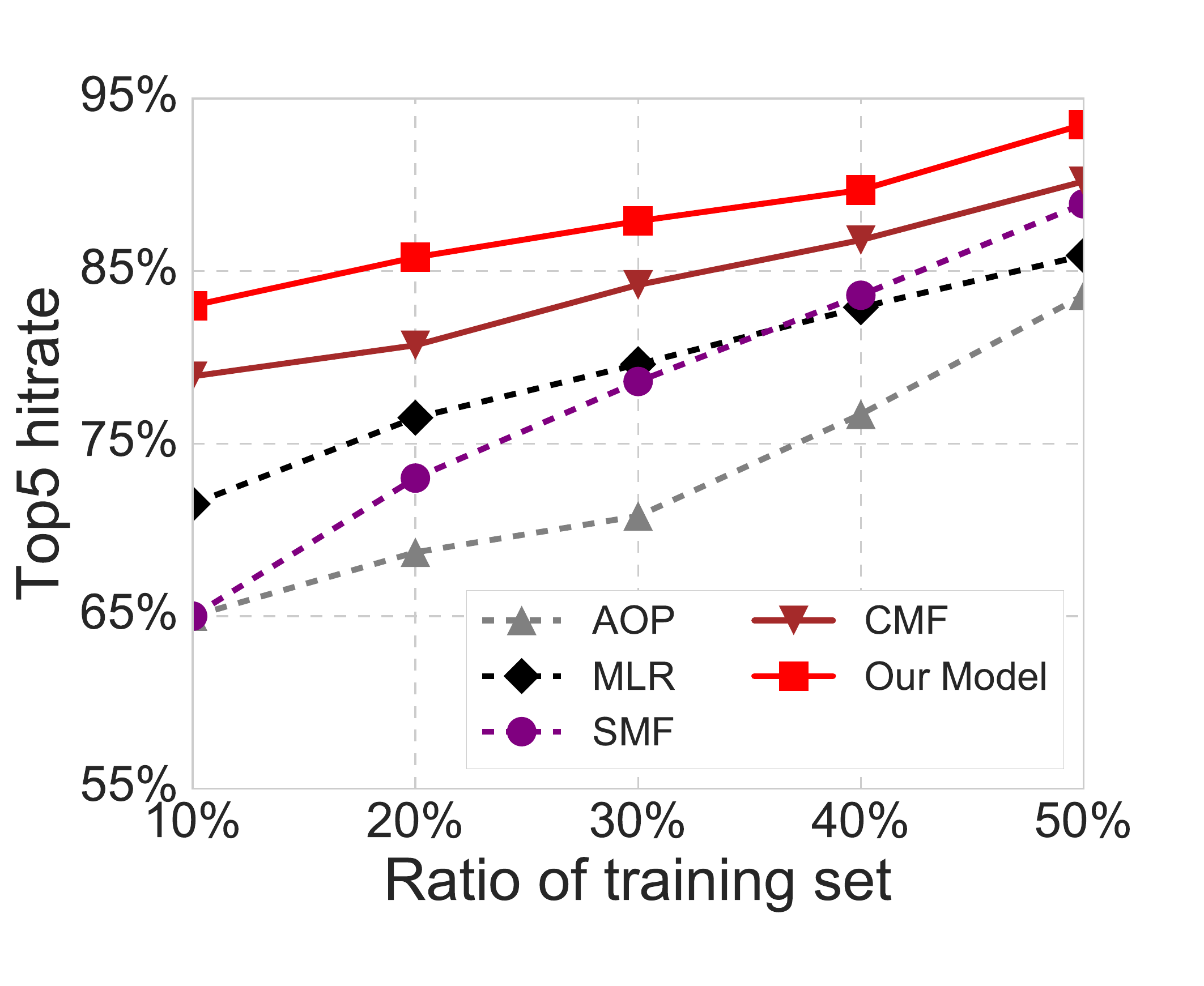}}
\subfigure[Top10 accuracy]{
\label{Fig.sub.2}
\includegraphics[width=5.1cm]{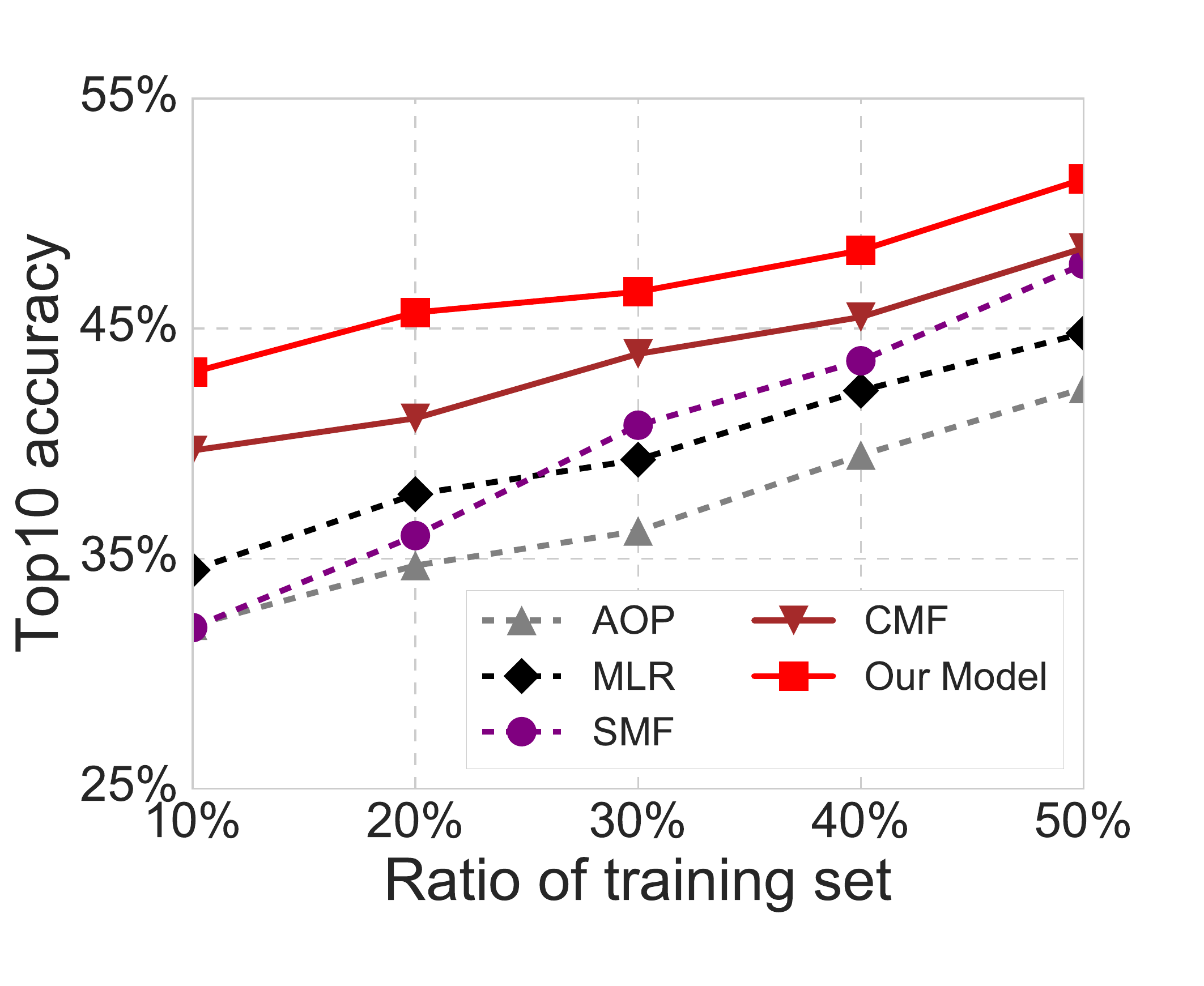}}
\subfigure[RMSE]{
\label{Fig.sub.3}
\includegraphics[width=5.1cm]{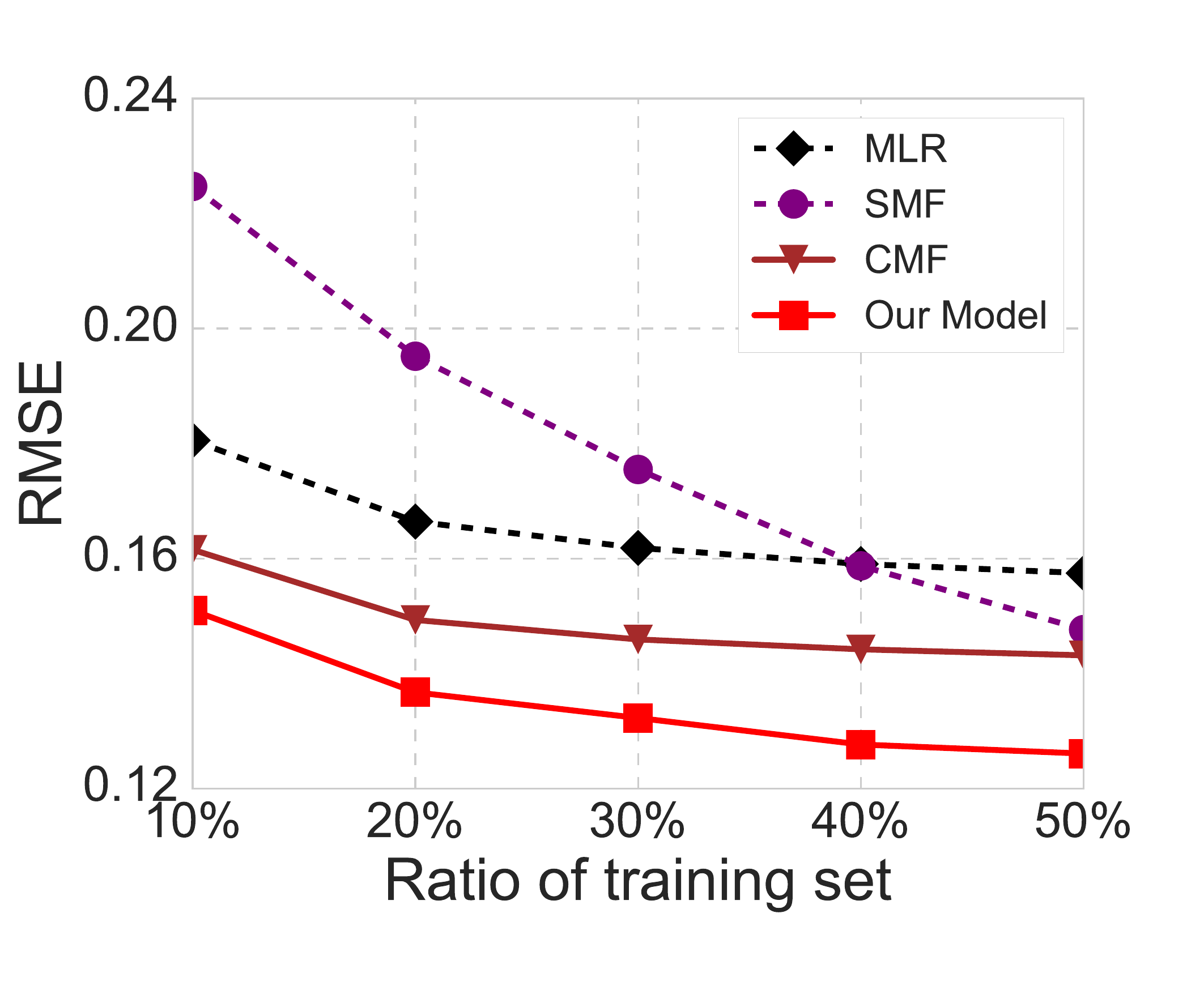}}
\caption{Performance evaluation for different sparsity levels.}
\label{Fig.label.sparsity}
\end{figure}

In this section we explore how our predictive model performs when the application usage data is at different sparsity levels. Specifically, we use different ratios of training data from $10\%$ to $50\%$ to test our algorithms. Training data $10\%$, for example, means we randomly select $10\%$ of the location-app data as the training data to predict the remaining $90\%$ of data. The results are shown in Fig.~\ref{Fig.label.sparsity}. 

Fig.~\ref{Fig.label.sparsity}(a) shows how top-5 hit rate changes under different sparsity levels for all compared the methods. More specifically, when sparsity levels vary from $10\%$ to $50\%$, our proposed model achieves a hitrate ranging between $83.0\%$ to $93.5\%$, which outperforms other methods with significant improvement. For example, with $20\%$ training data, our model improves by $6.3\%$ and $17.5\%$ compared to CMF and SMF respectively. Similarly, In Fig.~\ref{Fig.label.sparsity}(b), our model has the best top-10 prediction accuracy under different sparsity levels, with more than $43\%$ accuracy under $10\%$ training data. Fig.~\ref{Fig.label.sparsity}(c) indicates that our model also achieves lowest RMSE of $0.126$ $\sim$ $0.150$, which means that it achieves the best estimation about the overall usage distribution. These results suggest that our proposed model outperforms all the baseline methods. The performance gap between CMF and SMF increases as the training data becomes more sparse, which indicates that other information becomes more useful when there is not enough location-app data for training. The improvement between our model and CMF demonstrates that time-known factorization and transferring improve the prediction accuracy. In addition, we find that MLR does not perform well compared with CMF, which indicates that the locations with similar POI information does not necessarily have very similar app usage behaviors. Thus, regression with only POI information is not enough. In conclusion, POI and user information is useful for app usage prediction and our transfer learning model works best among the compared baseline method.

\subsection{Effect of User Number}
\begin{figure}[]
\centering
\subfigure[Top5 hitrate]{
\label{Fig.sub.1}
\includegraphics[width=7cm]{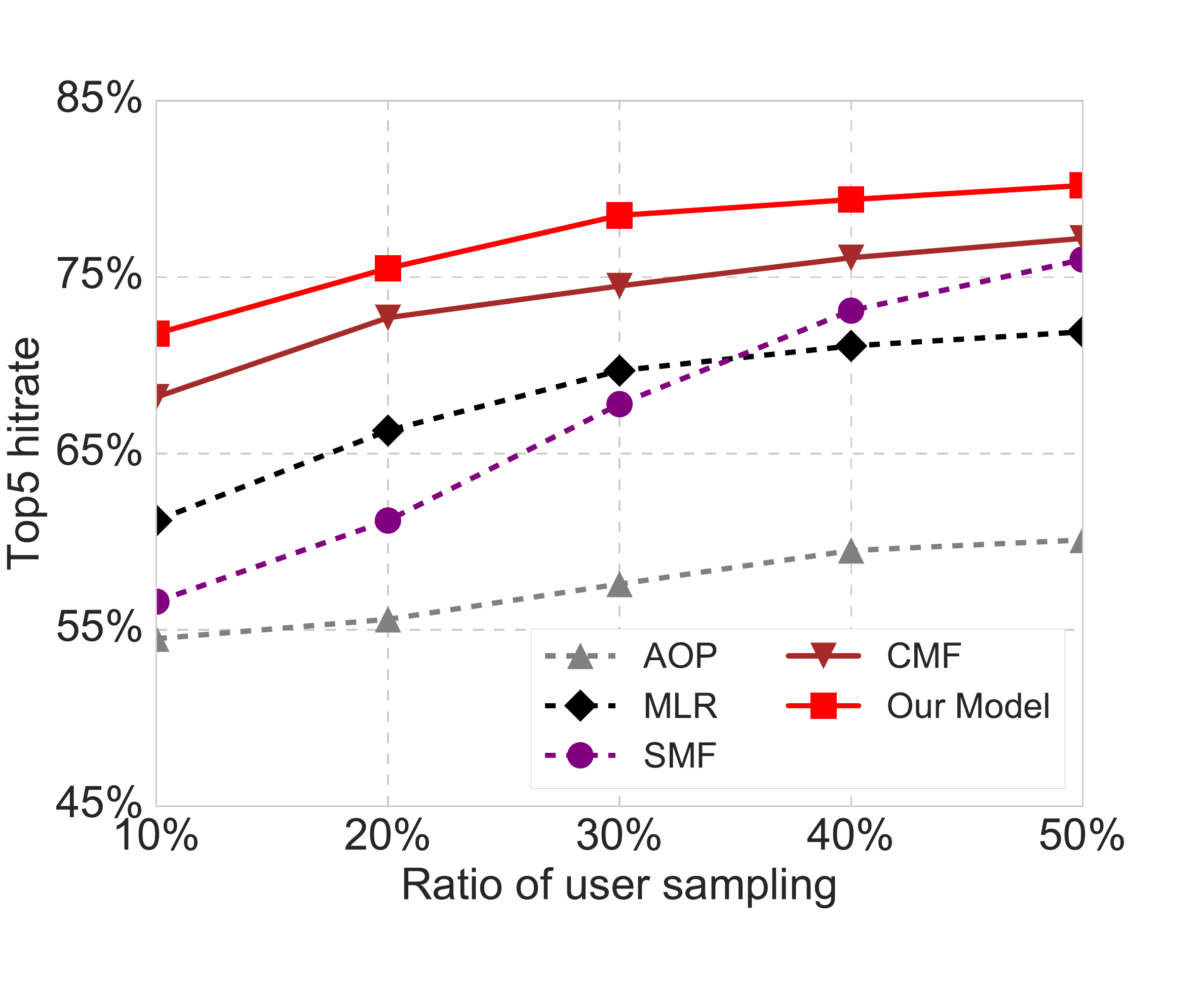}}
\subfigure[RMSE]{
\label{Fig.sub.2}
\includegraphics[width=7cm]{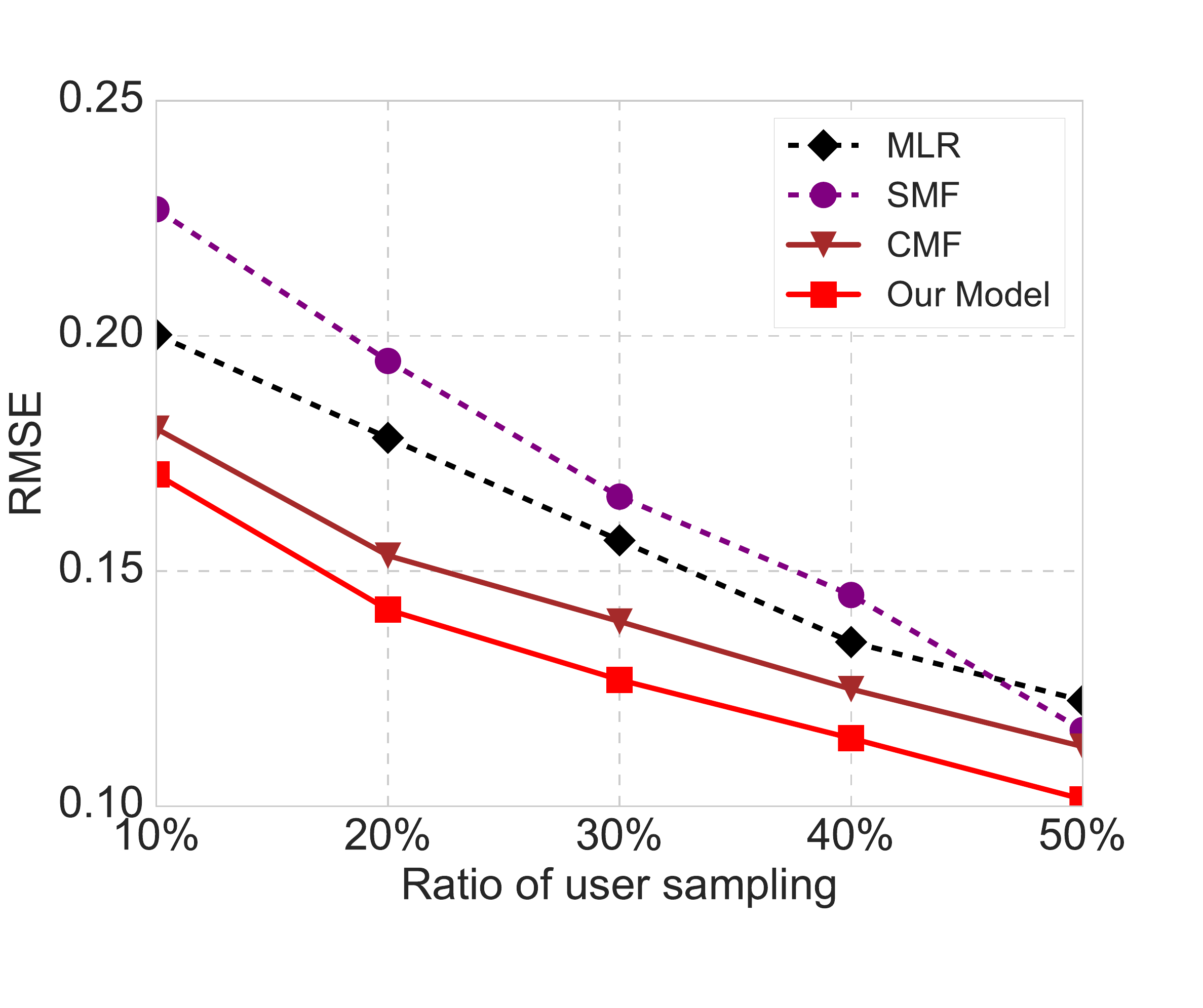}}
\caption{Performance evaluation for varying numbers of users.}
\label{Fig.label.users}
\end{figure}

As the mobile network becomes increasingly larger, we expect to have a growing number of mobile users. However, usually it is challenging to obtain the app usage records from all users, which means that the obtained app usage information in practice may have sampled bias from the truth value. We study the impact of the number of users so as to investigate the performance of our model under varying user sample sizes. Specifically, we sample users in varying ratios from $10\%$ to $50\%$ to test all models. For example, we randomly select $30\%$ of users to obtain the location-app matrix as the training data to predict all location-app data (contains $100\%$ users). Note that the location-correlation matrix will also be influenced by user-sampling. The results are shown in Fig.~\ref{Fig.label.users}. 

Fig.~\ref{Fig.label.users}(a) shows how top-5 hit rate changes under different user samples for all the compared models. Specifically, when the ratio varies from $10\%$ to $50\%$, our proposed model achieves hit rates from $71.8\%$ to $80.2\%$. Similarly, Fig.~\ref{Fig.label.users}(b) shows that our model achieves lowest RMSE, which is in the range of $0.101$ $\sim$ $0.170$. These results indicate that our proposed model performs better than other methods under sparse user sampling situation. Moreover, the gap between CMF and SMF increases as the number of users becomes smaller, when both location-app data and user information becomes insufficient. This suggests that POI information is very useful when we can only obtain data from a sample set of users.

\subsection{Effect of Spatial Resolution}

To validate the generalizability of our proposed model, we investigate the impact of spatial resolution. We vary the spatial resolution by merging neighboring locations into new larger units. Note that larger location size can protect the users' privacy better, since it will be more difficult to locate users accurately. In this analysis, we manipulate the spatial resolution from sector to base station and to street block. Typically one base station may contain two or three sectors, while a street block, which is the block part divided by streets, may contain three or four base stations. The number of sectors, base stations and street block is about 9800, 4800 and 1500 respectively. The average area of sector in the urban area and suburb area is about 0.24 km$^2$ and 0.95 km$^2$ respectively. Usually one sector contains over $80$ POIs, which means that the one sector will cover most of the typical types of POI. Thus, POI classification will be influenced lightly with the varying of space resolution.

\begin{figure}[]
\centering
\subfigure[Top5 hitrate]{
\label{Fig.sub.1}
\includegraphics[width=7.5cm,height = 5.5cm]{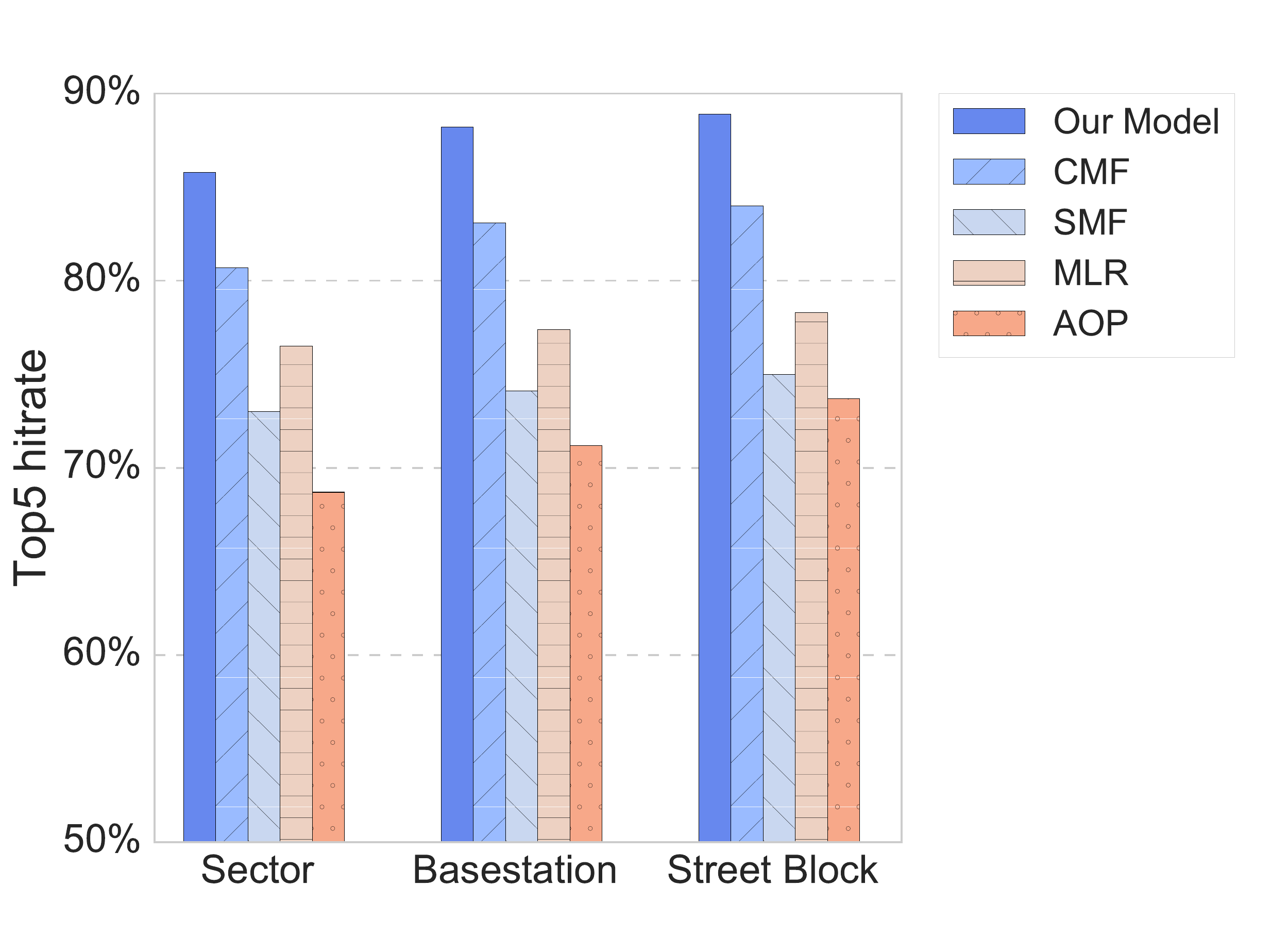}}
\subfigure[RMSE]{
\label{Fig.sub.2}
\includegraphics[width=7.5cm,height = 5.5cm]{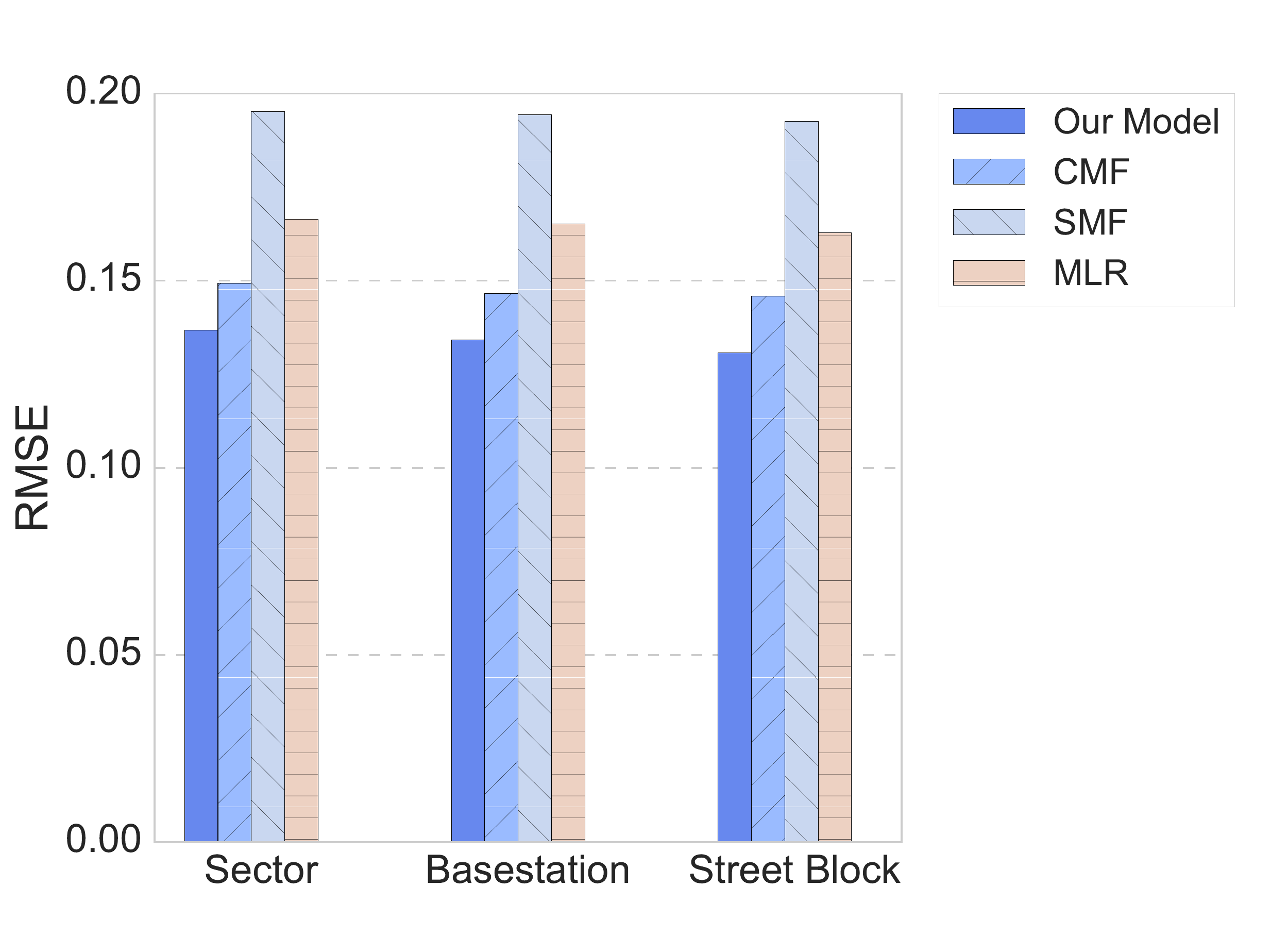}}
\caption{Performance evaluation for varying spatial resolution.}
\label{Fig.label.blocks}
\end{figure}

As previously, we randomly take $20\%$ of our data for training and $80\%$ as test data. The results are shown in Fig.~\ref{Fig.label.blocks}. From Fig.~\ref{Fig.label.blocks}(a) we observe that when the spatial resolution varies from sector to street block, our proposed model achieves best hit rate ranging from $85.8\%$ to $88.9\%$. Fig.~\ref{Fig.label.blocks}(b) shows that our model achieves lowest RMSE, which varies from $0.136$ to $0.130$. The results indicate that our proposed model outperforms the baseline models even under varying spatial resolution. Moreover, all models (including our own) tend to perform slightly better when the spatial resolution decreases (i.e. for larger spatial units). The reason is that the app usage tend to become more ``homogeneous" after location merging in a larger spatial areas.

\subsection{Cold-start Predictions}
We expect that in a realistic scenario we may not have app usage information from millions of users, but only have access to publicly available POI data. Thus, we investigate how our technique works in this case, which effectively resembles the cold start problem in recommender systems. In this scenario, SMF without other information cannot work since there is no location-app data at the test locations. To analyse the performance of our system, we randomly select $10\%$ of the locations for training, and use these to predict app usage for the remaining $90\%$ of locations. The results are shown in Table~\ref{tab:cold_start}. As expected, our proposed model outperforms the baselines and achieves $84.0\%$ hit rate and $0.148$ RMSE. Note that CMF performs only slightly better than MLR, because there is no usage data of any apps in the to-be-predicted locations. In conclusion, our model can also handle the cold-start problem.

\begin{table}[H]
\begin{center}
\begin{tabular}{c|c|c|c|c|c}
\hline
Metrics & Model & CMF & MLR  & AOP & SMF \\
\hline
Top5 hitrate & \textbf{0.84} & 0.77 & 0.74 & 0.57  & 0.04\\
\hline
RMSE & \textbf{0.148} & 0.167 & 0.172 & -  & 0.319\\
\hline
\end{tabular}\caption{Predictions for locations when prior data on app usage is not available.} \label{tab:cold_start}
\end{center}
\end{table}


\section{DISCUSSION AND RELATED WORK} \label{sec:Relatedwork}

A popular way for observing individuals with a smartphone has been through the recruitment of a sample of volunteers, and extrapolating that data onto a larger population. For instance, Work \& Tossavainen successfully transformed GPS traces from a few volunteers into a velocity field describing highway traffic \cite{4739016}. Similarly, Wirz et al. \cite{6269760} successfully estimated pedestrian movement and crowd densities at mass events using a subset of event attendees as probes who voluntarily shared their location using a mobile phone application. Their work suggests that tracking subsets of a crowd may provide enough information to reconstruct the movement of the whole crowd.

To track large scale pedestrian movement, a popular and scalable approach is in-network observation. For instance, Calabrese et al. \cite{5594641} estimated city-wide traffic by recording network bandwidth usage from signalling events, and showed how events taking place in the city can affect mobility patterns \cite{Calabrese2010}. With the popularity of location-based social networks \cite{zheng2011location}, users can share their real time activities by checking in at POIs, which provides a novel data source to study their collective behavior.  For example, Cheng et al. \cite{cheng2011exploring} investigated
22 million checkins across 220,000 users and reported a quantitative assessment of human mobility patterns by analyzing the spatial, temporal, social, and textual aspects associated with these footprints. Noulas et al. \cite{noulas2011empirical} conducted an empirical study of geographic
user activity patterns based on check-in data in Foursquare. Cranshaw et al. \cite{cranshaw2012livehoods} studied the dynamics of
a city based on user collective behavior in LBSNs. Wang et al. \cite{wang2014discovering} investigated the community detection
and profiling problem using users' collective behavior in LBSNs. Yang et al. \cite{yang2015nationtelescope} studied the large-scale collective behavior by introducing the NationTelescope platform to collect, analyzing and visualizing the user check-in behavior in LBSNs on a global scale. 

Since traditional LBSN can only get access to very limited mobile application data, a key contribution of our work is the collective app usage analysis based on POIs. To achieve this, we have investigated the types of applications that people use as they move around a city, particularly by considering the nearby POIs. A goal of our work is to analyze the statistical app usage at any given location which may contain hundreds of people, instead of the app usage of individual users. While a personalized prediction is suitable for studies where personal devices collect data, our dataset comes from the network side and provides data that is suitable for collective learning. Therefore, in our current study we did not consider factors related to individual app usage prediction. We also exclude app usage history as a factor, because the app usage prediction in our task is to predict the usage of app $j$ in location $i$ for the whole study period, which means that usage data remains unknown for the whole time. Considering historical usage data can be regarded as a time series prediction problem. Conversely, we do not know the usage history of app $j$ in location $i$. However, our model also incorporates the notion of "continuity" of time series by adding the loss function based on temporal continuity of latent feature vectors, which makes our model more accurate.

Our findings are very encouraging. Our results from Fig.~\ref{Fig.label.parameters} show that our model can predict with up to 85\% hit rate the top-5 popular applications at any given location across the city. In fact, these results remain robust when we vary the sparsity of our training approach (Fig.~\ref{Fig.label.sparsity}), the number of users (Fig.~\ref{Fig.label.users}) and the spatial resolution (Fig.~\ref{Fig.label.blocks}) of our analysis. Our Transfer Learning approach outperforms baseline approaches across all scenarios and parameters in our analysis. Ultimately, our work shows that it is possible to predict with higher accuracy which applications will be popular in a particular when we only consider the nearby POIs (Table ~\ref{tab:cold_start}). 

Understanding, and predicting, which types of applications people use is crucial and fundamental to a wide range of systems and operations, ranging from optimising battery life on the smartphone \cite{yan2012fast} to improving caching at the network to providing timely recommendations to users \cite{Shin:2012:UPM:2370216.2370243}. The importance is evident in the fact that this is already an extremely vibrant research topic in the UbiComp community and beyond, and additionally there is a rich literature on location-based services and recommenders which attempts to identify relevant services given a particular location.

\subsection{App Usage Prediction} 
A growing number of studies in recent years have sought to investigate the application usage on smartphones \cite{falaki2010diversity, xu2011identifying}. For example, detailed traces from 255 users are utilized to characterize smartphone usage from two intentional user activities: user interactions and application use \cite{falaki2010diversity}. Diverse usage patterns of smartphone apps are investigated via network measurements from a national level
tier-1 cellular network provider in the U.S \cite{xu2011identifying}. Jesdabodi et al. \cite{jesdabodi2015understanding} segmented usage data, which was collected from 24 iPhone users over one year, into 13 meaningful clusters that correspond to different usage states, in which users normally use their smarphone, e.g., socializing or consuming media. Jones et al. \cite{jones2015revisitation} identified three distinct clusters of users based on their app revisitation patterns, by analyzing three months of application launch logs from 165 users. Zhao et al. \cite{zhao2016discovering} analyzed one month of application usage from 106,762 Android users and discovered 382 distinct types of users based on their application usage behaviors, using their own two-step clustering and feature ranking selection approach. As a key step for mobile app usage analysis, i.e., classifying apps into some pre-defined categories, Zhu et al. \cite{zhu2012exploiting} proposed an approach to enrich the contextual information of mobile apps for better classification accuracy, by exploiting additional knowledge from a Web search engine.

However, most prior work mainly focuses on investigating app usage on individual level, and typically considers users' internal context. For instance, Huang et al. \cite{huang2012predicting} considers contextual information about last used application and time to predict the application that will be used next. The results showed that a regression model works best by incorporating identified sequences of application use in predicting the next application. This suggests a strong sequential nature in application usage on smartphones. Zhao et al. \cite{zhao2016prediction} proposed a method based on machine learning to predict
users' app usage behavior using several features of human mobility extracted from geo-spatial data in mobile Internet traces. Parate et al. \cite{parate2013practical} designed an app prediction algorithm, APPM, that requires no prior training, adapts to usage dynamics, predicts not only which app will be used next but also when it will be used, and provides high accuracy without requiring additional sensor context.

In fact, a lot of previous work has suggested that the applications people use are part of their behavioural habits, and are not necessarily linked to physical context. Considering routine, and focusing on overall mobile phone users' habits, Oulasvirta et al. \cite{oulasvirta2012habits} suggested that mobile phones are "habit-forming" devices, highlighting the "checking habit: brief, repetitive inspection of dynamic content quickly accessible on the device." This habit was found to comprise a large part of mobile phone usage, and follow-up work \cite{RN10721}  argued that the checking habit is one of the behavioral characteristics that leads to mobile application micro-usage, which is subsequently manifested as short bursts of interaction with applications.

\subsection{Location-aware Recommenders}
As wireless communication advances, research on location-based services using mobile devices has attracted interest. The CityVoyager system \cite{takeuchi2005outdoor} mines users' personal GPS trajectory data to determine their preferred shopping sites, and provides recommendations based on where the system predicts the user is likely to go in the future. Geo-measured friend-based collaborative filtering \cite{ye2010location} produces recommendations by using only ratings that are from a querying user's social-network friends that live in the same city. LARS \cite{levandoski2012lars} is a location-aware recommender system that uses location-based ratings to produce recommendations. It supports a taxonomy of three novel classes of location-based ratings, namely, spatial ratings for non-spatial items, nonspatial ratings for spatial items, and spatial ratings for spatial items. Yu et al. \cite{yu2012towards} proposed to mine user context logs (including location information) through topic models for personalized context-aware recommendation. Although these work consider location feature, they all focus on individual level recommendation.  

The spatial activity recommendation system CLAR \cite{zheng2010collaborative} mines the location data based on GPS and users' comments at various locations to detect interesting activities located in a city. It uses this data to answer two query types: (a) given an activity type, return where in the city this activity is happening, and (b) given an explicit spatial region, provide the activities available in this region. This is a vastly different problem than we study in this paper. CLAR focuses on five basic activities, but we want to estimate the location-based application usage data which is beneficial for location-based app recommendation containing thousands of apps. What's more, our data is collected from over 1 million users and locations are defined by base stations while CLAR only has 162 users and extract users' stay regions as locations by GPS trajectories data.

\subsection{How POIs Affect Our Behaviour}
Our results show that POIs have a strong effect on determining which applications are used near them. In fact, our analysis shows that we can predict with high accuracy the top-5 applications used at a given location by considering which POIs are nearby.

Our findings are supported by substantial literature that has investigated the effect of POIs, and in general land-use, on our behaviour in a variety of ways. A land-use approach has been often used in transportation research since the early 20th century. It describes the characteristics of travel behaviour between different types of land use, such as the traffic between residential zones and industrial zones. Voorhees \cite{Voorhees2013} described how travel between different types of origins and destinations roughly follows gravitational laws, with different types of destinations generating certain types of "pull" towards the origins. In fact, it is suggested that individuals organise spatial knowledge according to anchor points, POIs, or generally salient locations that form the cognitive map that the individual uses to navigate \cite{Manley2015123}. Besides geographical points, such as landmarks, anchor points can be path segments, nodes or even distinctive areas, similar to city properties categorized by Lynch \cite{lynch1960image}. McGowen et al. \cite{mcgowen2007evaluating} tested the feasibility of a model that predicts activity types based solely on GPS data from personal devices, GIS data and individual or household demographic data. Ye et al. \cite{ye2013s} proposed a framework which uses a mixed hidden Markov model to predict the category of user activity at the next step and then predict the most likely location given the estimated category distribution. Yang et al. \cite{yang2015modeling} first modelled
the spatial and temporal activity preference separately, and then
used a principle way to combine them for preference inference.

More broadly, land use effects various aspects of travel behaviour, such as trip generation, distance travelled and choice of mode of transport \cite{Boarnet2011}. Crucially, these effects seems to vary substantially according to the time of the day and week \cite{bromley2003disaggregating}. This provides us inspiration that consider separating the loss functions into different time periods, which increases the predictive accuracy of our model. At first this may appear as counter-intuitive since one might expect that larger data (and therefore longer periods) should yield the best results. However, as Bromley et al. \cite{bromley2003disaggregating} noted, the effects of POIs vary substantially during the day. As such, when narrowing the time period in our training data we effectively reduced the substantial variation of the POIs' effects, thus yielding better prediction results.

Finally, we should note that our work bears great resemblance to activity based models, which have often been used to estimate travel behaviour since the early 1990's \cite{axhausen1992activity}. Such models rely on the fact that people travel because they have needs and activities to which they must tend. How these activities are scheduled, given various conditions, such as household characteristics, properties of potential destinations and the state of the transportation network, is what activity based approaches seek to answer. However, activity based approaches have received criticism for their complexity and intense data requirements \cite{axhausen1998can}, and it has even been noted that it is difficult to find a representative set of participants willing to commit to a long-term data gathering effort \cite{Axhausen2002}. It is this exact weakness where our work can begin to make a contribution. Our work is the first to successfully bridge large-scale mobility data to large-scale activity data, albeit the latter is still at a rudimentary level of detail. Our work has sought to analyse application usage in terms of application "types" as grouped by appstores. However, it would also be possible to perform a more qualitative analysis of the role that applications play in users' everyday lives, and begin to map these to their urban mobility.   

\subsection{Limitations and Future Work}
Our work has a number of limitations. Our data was collected passively and anonymously, and therefore it is impossible for us to follow-up with participant questions and interviews to obtain qualitative data. In addition, our data is likely incomplete: only a subset of applications is captured through deep packet inspection. Applications that make no network requests were not captured in our dataset.

An important limitation of our dataset is that we are unable to tell the state of the application that makes the network requests. Specifically, it is not possible to distinguish between applications that made a network request after direct user input, and applications that run in the background and make network requests automatically. This ambiguity comes down to the definition of what does it mean to "use" an application. In our analysis, we assume that "use" means that the application exists on a user's phone, and is running. However, this definition does not imply that the user is explicitly interacting with the application.

Another limitation is that our dataset was collected over a period of one week. Although an entire population is captured in our dataset, it is well known that cities exhibit seasonal patterns which our dataset simply does not capture. These seasonal patterns may or may not affect the strength of our findings, but we can certainly expect that data from e.g. summer months may not be able to accurately predict behaviour during winter months.

About the future work, since the app usage information is obtained through networking analysis, it is worth to improve the prediction accuracy of app usage by utilizing the trace data that include how much traffic is generated from the app, in terms of number of packets or overall packet size. As for applications, further studies could be conducted to analyze and predict the performance for network operators, e.g., latency, throughput, or mobility management, etc., along with our app prediction system.

 \section{CONCLUSION} \label{sec:Conclusion}
In this paper we present, to the best of our knowledge, the first system to predict the Location-level app usage from the POI via a large-scale mobile data accessing records. Extensive evaluations and analysis reveal that our system outperforms three state-of-the-art methods in top-$N$ prediction accuracy and total app usage distribution estimation. We believe that our study provides a new angle to location-based app usage data mining, which paves the way for extensive applications including operating systems, network operators, appstores, profiling tools and advertisers.

\section*{Appendix 
\uppercase\expandafter{\romannumeral1}: Transfer Learning Model} \label{sec:Appendix I}

We use matrix factorization techniques to find a latent feature representation for locations, apps and POIs. What we transfer among the location-app domain, location-POI domain and the location-user domain is the latent feature of locations. We denote $L \in R^{K\times m}$, $A \in R^{K\times n}$ and $P \in R^{K\times l}$ to represent the latent location, app and POI matrices respectively, with column vectors $\textbf{l}_i$, $\textbf{a}_j$, $\textbf{p}_k$ representing the $K$-dimensional location-specific latent feature vector of location $i$, app-specific latent feature vector of app $j$, and POI-specific latent feature vector of POI $k$, respectively. For location latent feature vector $\textbf{l}_i$, we consider its components as functionality-based feature $\textbf{l}_i^1$ and user feature-based feature $\textbf{l}_i^2$, which means that $\textbf{l}_i = \textbf{l}_i^1 + \textbf{l}_i^2$. The corresponding location feature matrices are $L_1$ and $L_2$. 

We define the conditional distribution over the location-app matrix $X$, location-POI matrix $Y$ and the location correlation matrix $Z$ as follows, 

\begin{eqnarray*}
\begin{aligned}
\rho(X,Y,Z|L_1,L_2,A,P,\sigma_1^2,\sigma_2^2,\sigma_3^2) &=
\rho(X|L,A,\sigma_1^2) \rho(Y|L_1,P,\sigma_2^2)
\rho(Z|L_2,\sigma_3^2)\\
&= \prod N(x_{i,j}|g((\textbf{l}_i)^\top \textbf{a}_j),\sigma_1^2)  \times \prod N(y_{i,j}|g((\textbf{l}_i^1)^\top \textbf{p}_j),\sigma_2^2) \\
& \times \prod N(z_{i,j}|g((\textbf{l}_i^2)^\top \textbf{l}_j),\sigma_3^2)\ ,
\end{aligned}
\end{eqnarray*}

where $N(x|\mu, \sigma^2)$ is the probability density function of the Gaussian distribution with mean $\mu$ and variance $\sigma^2$. The function $g(x)$ is the logistic function $1/(1+exp(-x))$ to bound the range within $[0,1]$ interval, the same with our matrix data's range after preprocessing. From the conditional distribution above, we can observe that the latent feature vectors of locations are shared in both location-app domain and location-POI domain. We also place spherical Gaussian priors on location, app and POI feature vectors:

\begin{eqnarray*}
\begin{split}
& \rho(L_1|\sigma_{1,l}^2) = \prod_{i\in L}N(\textbf{l}_i^1|0, \sigma_{1,l}^2 I) , \
\rho(L_2|\sigma_{2,l}^2) = \prod_{i\in L}N(\textbf{l}_i^2|0, \sigma_{2,l}^2 I) , \\
& \rho(A|\sigma_a^2) = \prod_{j\in A}N(\textbf{a}_j|0, \sigma_a^2 I) , \
\rho(P|\sigma_p^2) = \prod_{k\in P}N(\textbf{p}_k|0, \sigma_p^2 I) \ .
\end{split}
\end{eqnarray*}

The generative process of our proposed model runs as follows:
\begin{itemize}
\item[$\bullet$] For each location $i$, draw the vector as $\textbf{l}_i = \textbf{l}_i^1 + \textbf{l}_i^2$ where $\textbf{l}_i^1 \sim N(0, \sigma_{1,l}^2I)$ and $\textbf{l}_i^2 \sim N(0, \sigma_{2,l}^2I)$,  
\item[$\bullet$] For each app $j$, draw the vector as $\textbf{a}_j \sim N(0, \sigma_a^2I)$,
\item[$\bullet$] For each POI $k$, draw the vector as $\textbf{p}_k \sim N(0, \sigma_p^2I)$,
\item[$\bullet$] For each location-app pair $(i,j)$, draw the value $x_{i,j} \sim N\left(g((\textbf{l}_i)^\top \textbf{a}_j),\sigma_1^2\right)$,
\item[$\bullet$] For each location-poi pair $(i,j)$, draw the value $y_{i,j} \sim N\left(g((\textbf{l}_i^1)^\top \textbf{p}_j),\sigma_2^2\right)$,
\item[$\bullet$] For each location-correlation pair $(i,j)$, draw the value $z_{i,j} \sim N\left(g((\textbf{l}_i^2)^\top \textbf{l}_j^2),\sigma_3^2\right)$.
\end{itemize}

Through Bayesian inference, the posterior probability of the latent feature vector sets $L$, $A$ and $P$ can be obtained as follows:

\begin{eqnarray*}
\begin{split}
\rho(L_1,L_2,A,& P|X,Y,Z, \sigma_1^2, \sigma_2^2,\sigma_3^2, \sigma_l^2, \sigma_a^2, \sigma_p^2) \\
& \propto \rho(X,Y,Z|L_1,L_2,A,P,\sigma_1^2,\sigma_2^2, \sigma_3^2)\rho(L_1|\sigma_{1,l}^2)\rho(L_2|\sigma_{2,l}^2)\rho(A|\sigma_a^2) \rho(P|\sigma_p^2) \\
& = \prod N(x_{i,j}|g((\textbf{l}_i)^\top \textbf{a}_j),\sigma_1^2) \prod N(y_{i,j}|g((\textbf{l}_i^1)^\top \textbf{p}_j),\sigma_2^2) \prod N(z_{i,j}|g((\textbf{l}_i^2)^\top \textbf{l}_j^2),\sigma_3^2)\\
& \times \prod_{i\in L}N(\textbf{l}_i^1|0, \sigma_{1,l}^2 I) \prod_{i\in L}N(\textbf{l}_i^2|0, \sigma_{2,l}^2 I) \prod_{j\in A}N(\textbf{a}_j|0, \sigma_a^2 I) \prod_{k\in P}N(\textbf{p}_k|0, \sigma_p^2 I)
\end{split}
\end{eqnarray*}  

The log of posterior distribution over the location, app and POI latent feature vector is calculated as:

\begin{eqnarray*}
\begin{split}
&\ln\rho(L_1,L_2,A,P|X,Y,Z, \sigma_1^2, \sigma_2^2,\sigma_3^2, \sigma_l^2, \sigma_a^2, \sigma_p^2)\\
& = -\frac{1}{2\sigma_1^2}\sum_{i,j} \left[ x_{i,j} - g((\textbf{l}_i)^\top \textbf{a}_j) \right]^2 -\frac{1}{2\sigma_2^2} \sum_{i,j} \left[ y_{i,j} - g((\textbf{l}_i^1)^\top \textbf{p}_j) \right]^2 
-\frac{1}{2\sigma_3^2} \sum_{i,j} \left[ z_{i,j} - g((\textbf{l}_i^2)^\top \textbf{l}_j^2) \right]^2 \\
& - \frac{1}{2\sigma_{1,l}^2}\sum_{i\in L_1}\Vert \textbf{l}_i^1 \Vert _2^2 - \frac{1}{2\sigma_{2,l}^2}\sum_{i\in L_2}\Vert \textbf{l}_i^2 \Vert _2^2 - \frac{1}{2\sigma_a^2}\sum_{j\in A}\Vert \textbf{a}_j \Vert _2^2   
 - \frac{1}{2\sigma_p^2}\sum_{k\in P}\Vert \textbf{p}_k \Vert _2^2  \\ 
 & - \frac{1}{2}  m(n\cdot \ln\sigma_1^2 + l\cdot \ln\sigma_2^2 + m\cdot \ln\sigma_3^2) - \frac{1}{2} K(m\cdot \ln\sigma_{1,l}^2\sigma_{2,l}^2 + n\cdot \ln\sigma_a^2 + l\cdot \ln\sigma_p^2)  + C \ ,
\end{split}
\end{eqnarray*}  

where $C$ is a constant that does not depend on the parameters. $\Vert \cdot \Vert_F^2$ denotes the Frobenius norm. Keeping the parameters, i.e., observation noise variance and prior variance, fixed, maximizing the log-posterior over the latent feature of locations, apps and POIs is equivalent to minimizing the following objective function, which is a sum of squared errors with quadratic regularization terms:

\begin{eqnarray*}
\begin{split}
\zeta (L_1,L_2,A,P)  = & \frac{1}{2}|| I \circ \left(X - g\left((L_1 + L_2)^\top A\right) \right)||_F^2 +\frac{\alpha}{2}|| Y - g\left(L_1^\top P\right) ||_F^2 +\frac{\beta}{2}||  Z - g\left(L_2^\top L_2\right) ||_F^2\\
& +  \left( 
\frac{\lambda_l^1}{2}||L_1||_F^2 +
\frac{\lambda_l^2}{2}||L_2||_F^2+
\frac{\lambda_a}{2}||A||_F^2 + \frac{\lambda_p}{2}||P||_F^2
\right) \ ,
\end{split}
\end{eqnarray*}

where $\circ$ means the point-wise matrix multiplication. $\alpha = \sigma_1^2/\sigma_2^2, \beta = \sigma_1^2/\sigma_3^2$ and $\lambda_l^1 = \sigma_1^2/\sigma_{1,l}^2$, $\lambda_l^2 = \sigma_1^2/\sigma_{2,l}^2$, $\lambda_a = \sigma_1^2/\sigma_a^2$, $\lambda_p = \sigma_1^2/\sigma_p^2$.

If taking time dynamics into consideration, for each time period $t$, the static location-POI matrix $Y$ does not change, and we share the same app latent feature $A$ at different time-specific loss functions to transfer knowledge among them since we consider the latent feature of apps keeps static, while the latent feature of POI is time-varying. Then we have the time-specific loss function as follows:

\begin{eqnarray*}
\begin{split}
\zeta (L_{1,t}.L_{2,t},A,P_t)  = &
 \frac{1}{2}|| I_t\circ \left(X_t - g\left((L_{1,t} + L_{2,t})^\top A\right) \right)||_F^2 +\frac{\alpha}{2}||  Y - g\left(L_{1,t}^\top P_t\right) ||_F^2 +\frac{\beta}{2}||  Z_t - g\left(L_{2,t}^\top L_{2,t}\right)||_F^2  \\
 &  + \left( 
\frac{\lambda_l^1}{2}||L_{1,t}||_F^2 +
\frac{\lambda_l^2}{2}||L_{2,t}||_F^2 +
\frac{\lambda_a}{2}||A||_F^2 + \frac{\lambda_p}{2}||P_t||_F^2
\right) 
\end{split}
\end{eqnarray*}

Finally, we consider the final loss function $\zeta$ is the sum of time-specific loss function of $\zeta (L_{1,t}.L_{2,t},A,P_t)$ during different time periods plus the loss function based on temporal continuity of time-specific latent feature vectors:

\begin{eqnarray*}
\begin{split}
\zeta = \sum_t \zeta (L_{1,t}.L_{2,t},A,P_t) + \left(\frac{\lambda_1}{2}  \sum_t\left(||L_{1,t} - L_{1,t-1}||_F^2 +
||L_{2,t} - L_{2,t-1}||_F^2\right)
 + \frac{\lambda_2}{2} \sum_t||P_t - P_{t-1}||_F^2 \right)
\end{split}
\end{eqnarray*}

Where the first term of the right part of equation means the sum of time-specific loss function and the second term stands for loss function based on temporal continuity.

Then, we perform gradient descent on $\textbf{l}_{i,t}^1,\textbf{l}_{i,t}^2, \textbf{a}_j, \textbf{p}_{k,t}$ for all locations, apps and POIs to get a local minimum of the objective function. The formulas run as follows:

\begin{eqnarray*}
\begin{split}
\frac{\partial \zeta}{\partial \textbf{l}_{i,t}^1} = & \sum_j \left[ g((\textbf{l}_{i,t})^\top \textbf{a}_j) - x_{i,j} \right]g'((\textbf{l}_{i,t})^\top \textbf{a}_j)\textbf{a}_j +
\alpha \sum_j \left[ g((\textbf{l}_{i,t}^1)^\top \textbf{p}_{t,k}) - y_{i,j} \right]g'((\textbf{l}_{i,t}^1)^\top \textbf{p}_{t,k})\textbf{p}_{t,k} \\
& + \lambda_l^1\textbf{l}_{i,t}^1 + \lambda_1[2\textbf{l}_{i,t}^1-(\textbf{l}_{i,t-1}^1 + \textbf{l}_{i,t+1}^1)] \ ; \\
\frac{\partial \zeta}{\partial \textbf{l}_{i,t}^2} = & \sum_j \left[ g((\textbf{l}_{i,t})^\top \textbf{a}_j) - x_{i,j} \right]g'((\textbf{l}_{i,t})^\top \textbf{a}_j)\textbf{a}_j +
\beta \sum_j \left[ g((\textbf{l}_{i,t}^2)^\top \textbf{l}_{j,t}^2) - z_{i,j} \right]g'((\textbf{l}_{i,t}^2)^\top \textbf{l}_{j,t}^2)\textbf{l}_{j,t}^2 \\
& + \lambda_l^2\textbf{l}_{i,t}^2 + \lambda_1[2\textbf{l}_{i,t}^2-(\textbf{l}_{i,t-1}^2 + \textbf{l}_{i,t+1}^2)] \ ; \\
\frac{\partial \zeta}{\partial \textbf{a}_j} = & \sum_t \left( \sum_i \left[ g((\textbf{l}_{i,t})^\top \textbf{a}_j) - x_{i,j} \right]g'((\textbf{l}_{i,t})^\top \textbf{a}_j)\textbf{l}_{i,t} + \lambda_a\textbf{a}_j \right) \ ; \\
\frac{\partial \zeta}{\partial \textbf{p}_{k,t}} = & \alpha \sum_i \left[ g((\textbf{l}_{i,t}^1)^\top \textbf{p}_{k,t}) - y_{i,k} \right]g'((\textbf{l}_{i,t}^1)^\top \textbf{p}_k)\textbf{l}_{i,t}^1 + \lambda_p\textbf{p}_{k,t} + \lambda_2[2\textbf{p}_{k,t}-(\textbf{p}_{k,t-1} + \textbf{p}_{k,t+1})] \ ,
\end{split}
\end{eqnarray*}   

where $g'(x)$ is the derivative of the logistic function and $g'(x) = exp(-x)/(1+exp(-x))^2$.

There exist several methods to reduce the time complexity of model training, and we adopted mini-batch gradient descent approach to learn the parameters. With random sampling, the cost of the gradient update no longer grows linearly in the number of entities related to latent feature vectors, but only in the number of entities sampled. The hyper-parameters, i.e., number of latent features and regularization coefficient, are set by cross-validation.

\section*{APPENDIX \uppercase\expandafter{\romannumeral2}: Evaluation Metrics} \label{sec:Appendix II}
For each location $i$ in a test set, we predict the usage data $\widehat{x}_{ij}$ for each app in the candidate set, where $\widehat{x}_{ij}$ is estimated as $(\textbf{l}_i)^\top \textbf{a}_j$ with different approaches to learn $\textbf{l}_i$ and $\textbf{a}_j$. Top N prediction list is obtained by sorting $\widehat{x}_{ij}$ in a descending order and keeping the first N apps. For the AOP model, the apps in the candidate set are sorted according to the total usage amount in the training data and we cannot get the specific usage data. Suppose there are $L_{test}$ locations in the test data, and for location $i$, $V_i^{test}$ and $V_i^{pre}$ stand for the real top-$N$ apps set and the predicted top-$N$ apps set respectively,  while $\textbf{r}_i^{test}$ and $\textbf{r}_i^{pre}$ are the real and predicted usage data of apps in location $i$'s candidate set respectively.

In recommendation systems, Top-$N$ hit rate, which is the percentage of locations whose top-$N$ apps are successfully predicted (correct for at least one app), is commonly since they usually recommend a list of apps to expect users click at least one of them. The accurate calculation runs as follows:
\begin{eqnarray*}
\begin{split}
TopN-{\rm hitrate} = \left( \sum_i \left( |V_i^{test} \cap V_i^{pre}| \geq 1 \right) \right) / L_{test} \ .
\end{split}
\end{eqnarray*}

We also use Top-$N$ prediction accuracy, which stands for the mean prediction accuracy on top-$N$ prediction of all locations, as performance evaluation metrics:
\begin{eqnarray*}
\begin{split}
TopN-{\rm accuracy} = \left( \sum_i \frac{|V_i^{test} \cap V_i^{pre}|}{N} \right) / L_{test} \ .
\end{split}
\end{eqnarray*}

The above two metrics mainly focus on the popular apps. However, the overall usage distribution may also be important for some potential applications. In order to measure the overall distribution, we use RMSE to measure the error between the true and estimated app usage, which is defined as follows:
\begin{eqnarray*}
\begin{split}
RMSE = \left( \sum_i \Vert \textbf{r}_i^{test} -  \textbf{r}_i^{pre} \Vert_2 \right) / \sum_i len(\textbf{r}_i^{test}) \ .
\end{split}
\end{eqnarray*}

\bibliographystyle{unsrt}
\bibliography{refer}

\end{document}